\documentclass[12pt,preprint,xdvi]{aastex}

\usepackage{natbib}
\usepackage{graphicx}

\newcommand{\be}{\begin{equation}}
\newcommand{\ee}{\end{equation}}
\newcommand\eq{eq.}

\newcommand\fig{Fig.}
\newcommand\figs{Figs.}

\def\vvec{{\bf v}}

\def\rvec{{\bf r}}
\def\gvec{{\bf g}}
\def\xhat{{\bf \hat{x}}}

\def\zhat{{\bf \hat{z}}}

\def\lhat{{\bf \hat{l}}}

\def\half{\hbox{${1\over2}$}}

\def\paperI{LDCQ}

\def\cv{{c}_{_{\rm V}}}

\def\kb{k_{_{\rm B}}}

\begin{document}

\title{A reconnection-driven model of the hard X-ray loop-top source from flare 2004-Feb-26}

\author{Dana Longcope$^1$, Jiong Qiu$^1$, and Jasmine Brewer$^{1,2}$}
\affil{1. Department of Physics, Montana State University,
  Bozeman, Montana USA 59717\\
  2. University of Colorado at Boulder, Boulder, CO, USA 80303 }




\begin{abstract}
A compact X-class flare on 2004-Feb-26 showed a concentrated source of hard X-rays at the tops of the flare's loops.  This was analyzed in previous work (Longcope {\em et al.} 2010), and interpreted as plasma heated and compressed by slow magnetosonic shocks generated during post-reconnection retraction of the flux.  That work used analytic expressions from a thin flux tube (TFT) model, which neglected many potentially important factors such as thermal conduction and chromospheric evaporation.  Here we use a numerical solution of the TFT equations to produce a more comprehensive and accurate model of the same flare, including those effects previously omitted.  These simulations corroborate the prior hypothesis that slow mode shocks persist well after the retraction has ended, thus producing a compact, loop-top source instead of an elongated jet, as steady reconnection models predict.  Thermal conduction leads to densities higher than analytic estimates had predicted, and evaporation enhances the density still higher, but at lower temperatures.  X-ray light curves and spectra are synthesized by convolving the results from a single TFT simulation with the rate at which flux is reconnected, as measured through motion of flare ribbons, for example.  These agree well with light curves observed by RHESSI and GOES and spectra from RHESSI.  An image created from a superposition of TFT model runs resembles one produced from RHESSI observations.  This suggests that the HXR loop-top source, at least the one observed in this flare, could be the result of slow magnetosonic shocks produced in fast reconnection models like Petschek's.
\end{abstract}

\section{Introduction}

Hard X-ray (HXR) imaging observations of solar flares frequently show sources of concentrated emission from one or both footpoints of the flaring loop.  Such footpoint sources are believed to arise from the deposition of the flare energy in the denser chromosphere, probably carried through the low-density corona by non-thermal electrons.  Less frequently there is a source located between the footpoints in the middle of the flaring loop or slightly above it \citep{Masuda1994,Petrosian2002,Krucker2008b}.  Such loop-top sources are generally fainter than typical footpoint sources and are therefore most often observed when the footpoints are either occulted, or lack sources of their own.  In the latter cases the electron spectrum is either thermal at very high temperature  \citep[$\ga 30$ MK, termed {\em super-hot,}][]{Lin1981,Lin1985,paperI,Caspi2010,Caspi2014} or have non-thermal spectra with large power-law index 
\citep[i.e.\ a soft spectrum,][]{Veronig2004,Veronig2005}.  In either of these cases the intense coronal emission and lack of footpoint sources are generally assumed to require very 
high coronal density, $n_e\ga 10^{11}\,{\rm cm}^{-3}$.  The explanation for such sources remains controversial and will be the subject of this investigation.

Loop-top or above-the-loop-top HXR sources are generally believed to be associated with the coronal energy release which is powering the flare.  It is not clear, however, how this energy release produces the high densities required, and then how it manages to confine that density and pressure for the minutes the sources typically last.  Under one hypothesis the sources occur at magnetic null points where particles are both accelerated and confined \citep{Somov1997,Karlicky2004,Veronig2006}.  The main challenge faced by this explanation is for a collisionless trap to work at densities large enough to produce the observed emission and to thermalize such an energetic electron population.  \citet{Jiang2006} proposed that plasma turbulence has reduced the field-aligned thermal conductivity enough for the source to remain hot, localized, and persistent beyond the classical cooling time.  While providing temperature localization, the suppression of thermal conduction will not supply a force to confine the pressure concentrated at the loop top.

Considering one particular flare with no footpoint sources, \citet{Veronig2004} proposed that chromospheric evaporation had raised the loop's density enough (they estimate $n_e\simeq2\times10^{11}$) to prevent accelerated coronal electrons from reaching the footpoints.  Since evaporation is generally considered responsible for the high densities observed throughout a flare, it does seem a natural explanation for the high-density loop-top source.  Invoking it in this role, however, poses a challenge of timing.  
Magnetic energy must first be released in the corona and propagate to the chromosphere to drive evaporation.  The evaporated material must then make its way back to the corona to affect 
the energy release process responsible for its own creation.
Microwave observations of this flare revealed pre-flare densities as much as one-quarter as high \citep{Veronig2005}, but that still leaves evaporation to supply the remaining three-quarters during the flare itself.  At the very least this would require that the energy release remain active on a single field line a longer time than evaporation takes to reach the loop's apex.

\citet{Sui2003} and \citet{Sui2004} observed a pair of HXR sources above loops in a series eruptive limb flares over the period 2002-Apr-14 -- Apr-16.  Based on their positioning and evolution these were determined to be associated with the current sheet (CS) in which reconnection was occurring.  Magnetic reconnection is widely believed to be the source of energy in flares, and the mechanism most likely to work in a flare is that proposed be Petschek's \citep{Petschek1964}.  Petschek's model requires that reconnection occur within a small portion of a larger CS, with outflow jets approaching the Alfv\'en speed moving plasma and magnetic flux through the CS.  The structure and evolution observed by \citet{Sui2003} and \citet{Sui2004} seem consistent with such a mechanism.
The detailed physics producing the flux transfer within the CS which initiates the energy release is still a matter of investigation.  Recent proposals have been for kinetic effects \citep{Shay1998,Rogers2001,Hesse2004,Fujimoto2006}, plasmoid instabilities \citep{Loureiro2007,Bhattacharjee2009,Pucci2014}, or turbulence within the sheet \citep{Lazarian1999,Higashimori2013,Huang2016}.  Regardless of how it is initiated, 
it is clear that energy is released to produce the flare effects we observe.

Modeling by \citet{Tsuneta1998} suggested that loop-top sources might occur at the ends of a CS undergoing Petschek reconnection where fast magnetosonic termination shocks (FMTSs) are expected to form \citep{Forbes1986c}.  Numerical simulations have predicted FMTSs to occur, at least in certain cases, at the end of the current sheet where external plasma and fields stop the reconnection outflow jet.  The density within the FMTS depends on the flow speed relative to the fast magnetosonic speed, but most simulations show an enhancement of about 50\% or so above that in the outflow jet \citep{Forbes1986c}.  With such a modest density, and the relatively small volume occupied by the shock itself, it is problematic for FMTSs to explain the significant densities and emission measures of the observed loop-top sources.

\citet{paperI} and later \citet{Longcope2011b} suggested that loop-top sources were the manifestations of a different feature more essential to Petschek's model: the outflow jets and {\em slow magnetosonic shocks} (SMSs).  In the basic Petschek model, magnetic energy is released through the retraction of reconnected field lines, driving plasma outflow at close to the local Alfv\'en speed.  In a two-dimensional, steady-state model, this outflow takes the form of a narrow, steady jet originating at the reconnection site \citep{Petschek1964,Vasyliunas1975,Soward1982b,Priest1986}.  In more recent variations \citep{Longcope2009,Guidoni2010,Guidoni2011,Longcope2015b}, reconnection can occur more sporadically, and the outflow can be less coherent, perhaps forming a disordered collection of flux tubes (or ``plasmoids'') retracting though the current sheet.  In either event, a portion of the flow's kinetic energy is thermalized in SMSs.    These shocks heat the plasma at the same time they compress it by up to a factor of 20 \citep{Longcope2011b}.  In unsteady versions the SMSs travel with the retracting tube, bounding a moving plug of hot compressed plasma, rather than a standing wedge.  Moreover, the compressive flows initiated during retraction are not necessarily stopped when the retraction stops, and can continue driving SMSs while the tube rests in the post-flare arcade \citep{paperI,Longcope2015b}.  This offers the possibility that Petschek-like SMSs have sufficient temperature and emission measure, and the correct location, to account for the observed loop-top sources.

\citet[hereafter called \paperI]{paperI} attempted to test this hypothesis using the loop-top source observed in the X-class disk flare of 2004-Feb-26 (SOL2004-02-26T02:03:00).  RHESSI observations of this flare reveal spectra with little evidence of non-thermal electrons at the peak time or afterwards.  Thermal fits to the spectra show a super-hot component, $T\ga30$ MK, with $EM\simeq 4\times10^{48}\,{\rm cm^{-3}}$.  Images of the 12--25 keV emission show a concentration centered around the place where post-flare loops seem to have their apices -- i.e.\ at the loop-tops.  (Since the flare occurred on the solar disk it was not possible to determine if the emission was on or {\em above} these loops.  We therefore use the term loop-top source to include the possibility that it lies above lower-temperature loops.)

The flare observations were modeled using an unsteady version of Petschek reconnection, called the thin flux tube model \cite[TFT,][]{Linton2006,Longcope2009,Guidoni2010,Guidoni2011,Longcope2015b}.  \citet{paperI} used analytic expressions from this model, based on Rankine-Hugoniot (RH) conditions of the shocks, to estimate the temperature and density of the outflow.  They went on to derive analytic estimates for the size and duration of that feature, predicting its survival well beyond the end of the retraction phase.  They were not able, however, to account for possible effects of chromospheric evaporation on the retraction-generated source.  

Recent numerical solutions of the TFT model have revealed some potential shortcomings in the analytical approach used in \paperI.  Some simulations have revealed that thermal conduction renders the RH conditions inapplicable to SMSs in solar flares \citep{Longcope2011b}.  Others have shown that chromospheric evaporation can interact strongly with the shocks \citep{Longcope2015b}.  The present work returns to the task of modeling the loop-top source of the 2004-Feb-26 flare, but this time using numerical solutions of the TFT model.  We develop here a convolution technique to obtain synthetic observations from a single-loop simulation.  The results so obtained corroborate the viability of the SMS as the source of loop-top sources observed in HXR.

This new analysis will be presented as follows.  The next section reviews the observations of 2004-Feb-26, and presents some updated analyses.  Section 3 reviews the TFT model and explains how observations are used to constrain two of the five loop parameters.  It then presents the simulation of a single loop, taken to represent all loops created by reconnection in the flare.  A novel technique is presented in section 4 for using convolution to synthesize light curves.  This is applied to the simulation and the observed reconnection rate to obtain light curves comparable to those from observations.  In section 5, the single-loop simulation is used to synthesize an image in the 12--25 keV energy band, comparable to an image formed from RHESSI data from the flare.  Finally, section 6 discusses the significance of the results.

\section{Flare 2004-Feb-26}

\begin{figure}[htbp]
\centerline{\includegraphics[width=7.0in]{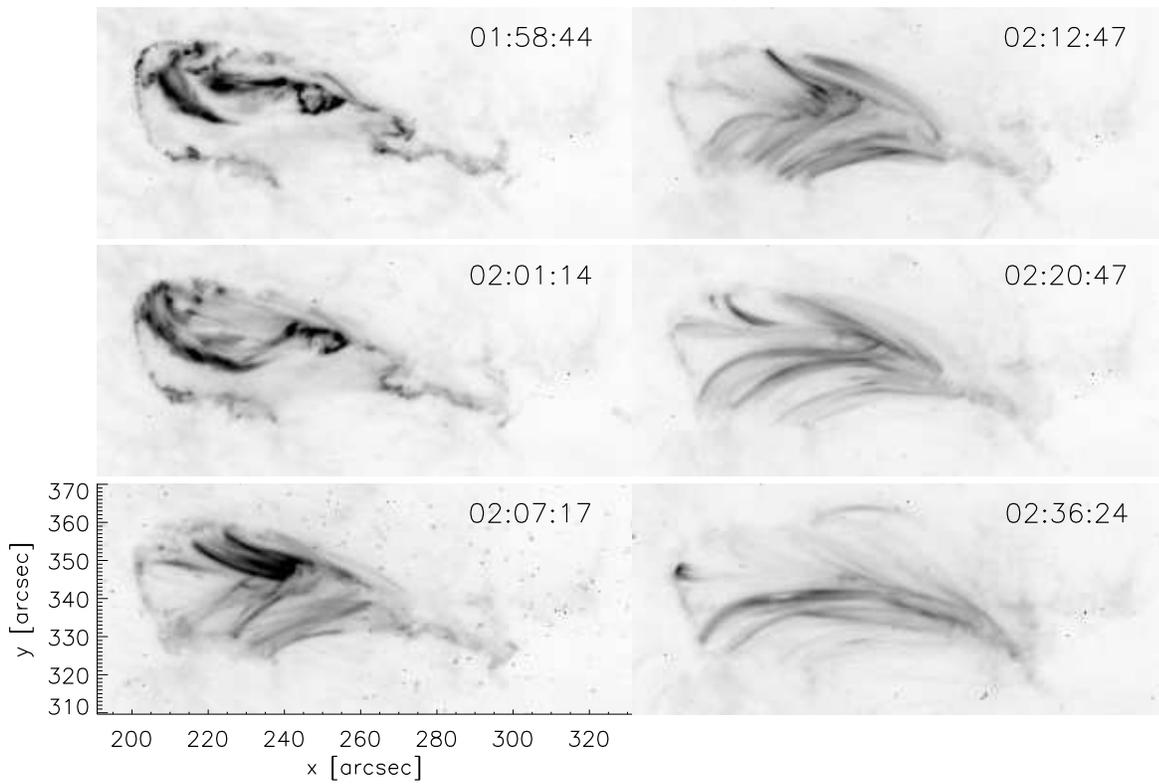}}
\caption{Selected images extracted from TRACE 171 \AA\ data plotted as an inverse square-root grey scale.}
	\label{fig:trace}
\end{figure}

Our flare occurred at 2:00 on 2004-Feb-26 in AR10564 ($14^{\circ}$N) when it was $14^{\circ}$ West of the central meridian.  The active region (AR) was basically bipolar, but had experienced about 50 hours of additional flux emergence prior to the flare.  TRACE recorded 171 \AA\ images of the AR at 30-second cadence throughout the flare.  Several images extracted from this data sequence are shown in \fig\ \ref{fig:trace}.  Images from the later phases show numerous post-flare loops which have cooled into the 171 \AA\ passband at $T\simeq10^6$ K  (see \fig\ \ref{fig:trace}).  \paperI\ performed a detailed analysis of these loops, identifying 143 different loops, some of which remained visible beyond 3:20.  We take these distinct loops to be the product of an unsteady, fast reconnection process responsible for the flare.

\begin{figure}[htbp]
\centerline{\includegraphics[width=6.0in]{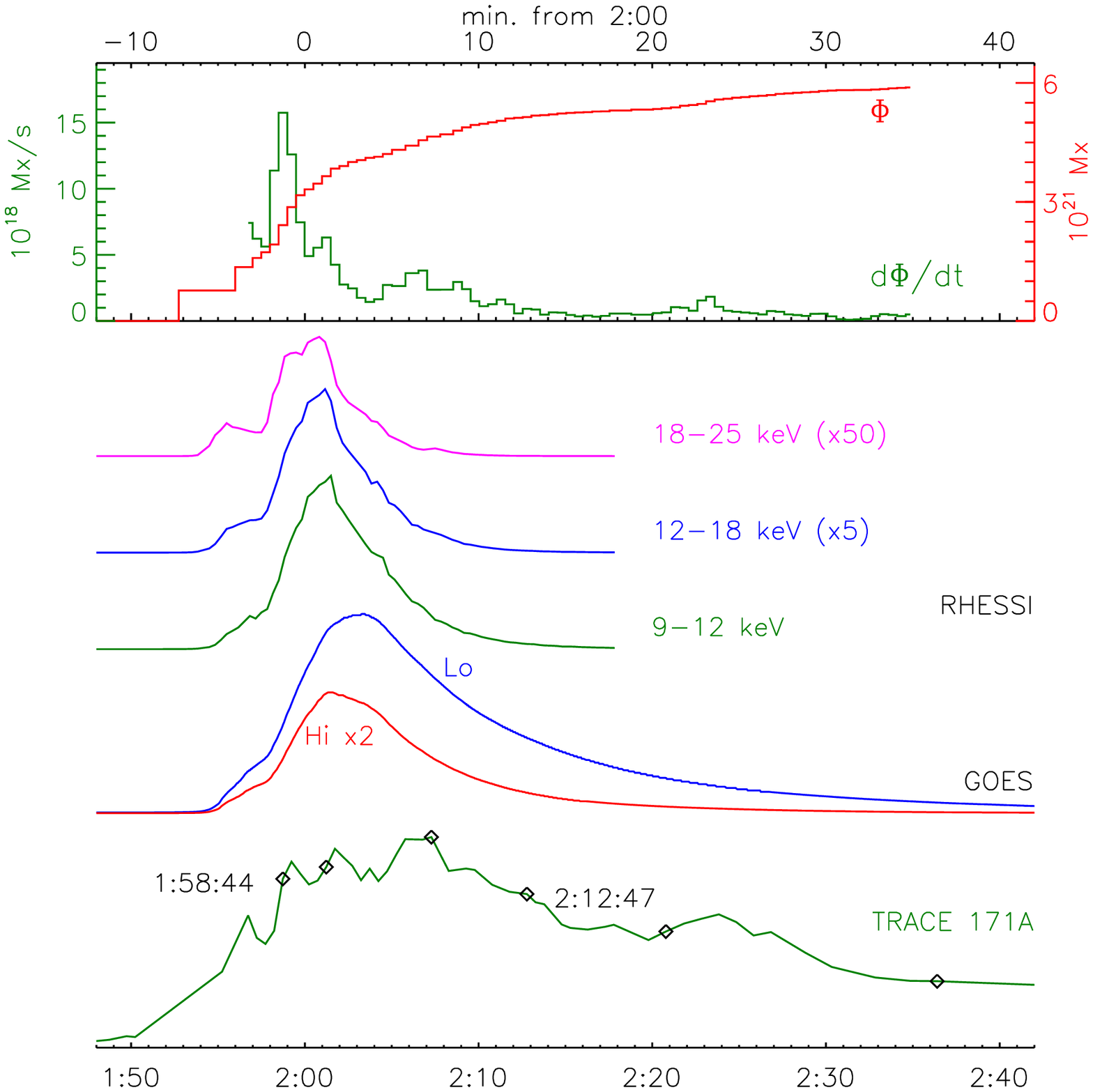}}
\caption{Light curves summarizing the flare on 2004-Feb-26.  From bottom to top are plotted, integrated TRACE 171 \AA\ emission (green), GOES high (red) and low (blue) energy band, and RHESSI curves from $9$--$12$ keV (green), $12$--$18$ keV (blue), and  $18$--$25$ keV (red).  All are on linear scales, with scales adjusted and displaced vertically for clarity.  Diamonds on the TRACE curve correspond to the images from \fig\ \ref{fig:trace}.  The top panel shows the net flux swept by ribbons (red, against right axis), and its derivative 
$\dot{\Phi}_{\rm rib}$ (green against left axis).}
	\label{fig:thist}
\end{figure}

RHESSI was observing from well before the flare (1:35) until a time during the decay phase (2:30), and thus observed the entirety of the flare's impulsive phase.  Light curves from three representative energy bands are plotted in \fig\ \ref{fig:thist} along with the summed counts from TRACE 171\AA\ (green) and GOES (red and blue).  The highest-energy bands (RHESSI and GOES high) show similar profiles peaking around 2:01:30.  An early bump (1:53--1:57)  in the 18--25 keV curve (magenta) is evidence of the only period where the HXR spectra showed non-thermal particles.  By the onset of the main rise the photon spectrum is best fit by contributions from two Maxwellian distributions with no power law component (\paperI).

The RHESSI data used by \paperI\ has been reanalyzed for the present work. The observed data counts are converted to a time-dependent photon spectrum, in units of photons/s/cm$^2$/keV, to compare with model-derived spectra and facilitate the adjustment of model parameters. For the conversion, the data from detector 4 is used and integrated over 20 second intervals.  As in \paperI, we use a fitting model with two thermal components to achieve the best match with observed counts spectrum.  During the main phase of the flare from 1:55 to 2:10 UT, RHESSI data are taken mostly with attenuator number 3 and occasionally with attenuator 1, and the range of the data for fitting is from photon energy 6~keV. The reduced chi-square of the fitting during this period is about 1, indicating good convergence. The RHESSI light curves in \fig\ \ref{fig:thist} are found by integrating the photon spectra over the ranges of photon energies indicated.

The process of magnetic reconnection is best characterized and quantified using high-cadence images of the flare ribbon made in a chromospheric spectral line \citep{Qiu2002,Qiu2009}.  While no such observations were made of this flare, it is possible to identify the ribbons in the 171 \AA\ images (see \fig\ \ref{fig:trace}).  These ribbons map out the reconnection which produces the flux tubes we see later as coronal loops in the same data.  Superposing the ribbon locations on the pre-flare MDI line-of-sight magnetogram allows a calculation of the flux $\Phi_{\rm rib}(t)$ swept up by the ribbon over time.   This calculation was performed in \paperI, and is repeated in this investigation with some modifications. We now use a lower threshold to pick up weaker ribbon emission especially in the positive ribbon. We also use the photospheric magnetogram in this study, but do not extrapolate it to 2000 km above the photosphere, as we had previously done. Finally, \paperI\ measured the reconnection flux from only the negative ribbon, while in this study we compute the reconnected flux from the average of the measurements from both ribbons. These changes give rise to greater reconnection flux and reconnection rate than those in \paperI.  The result, shown along the top panel of \fig\ \ref{fig:thist} records a total reconnected flux of $\Delta\Phi_{\rm rib}\simeq6\times10^{21}$ Mx.  The reconnection rate peaks at $\dot{\Phi}_{\rm rib}=1.6\times10^{19}\,{\rm Mx/s}$ at 1:58:50, just before the X-ray fluxes peak.

\section{The Flare loop model: TFT}

Our principal hypothesis is that the flare was caused by the rapid release of magnetic energy through reconnection.  We model the energy conversion from this reconnection using the TFT model \citep{Linton2006,Longcope2009,Guidoni2010,Guidoni2011,Longcope2015b}.  Like Petschek's original model, TFT considers the plasma dynamics {\em after} magnetic reconnection has occurred across a CS.  Because energy is released through shortening of the reconnected field lines following the actual reconnection, the TFT model does not include the reconnection itself, but only the dynamics of the energy release as the reconnected field lines retract through the current sheet.  We believe it is these dynamics which produce all the observable effects of the solar flare.

We use the simplest version of the TFT model in which the CS separates uniform layers of magnetic flux of equal magnitude, $B$, but differing in direction by an angle $\Delta\theta$, called the shear angle.  Reconnection between two straight flux tubes from opposite sides of the CS, will form a pair of new flux tubes bent at the point of their joining by an interior angle 
$180^{\circ}-\Delta\theta$.  We model lower flux tube whose subsequent retraction adds to the observed post-flare arcade, and later appears as a loop in TRACE 171 \AA.  This initially bent tube, with initial length $L_0$, releases magnetic energy by retracting to a final length $L_{\rm fin}=L_0-\Delta L$.  A tube of flux $\delta\psi$ undergoing this retraction will release energy equalling $\delta\psi B\Delta L/4\pi$ \citep{Longcope2011b}.  Under the reconnection hypothesis, this is the energy powering the flare.

\subsection{The loop properties}

\begin{figure}[htbp]
\centerline{\includegraphics[width=6.0in]{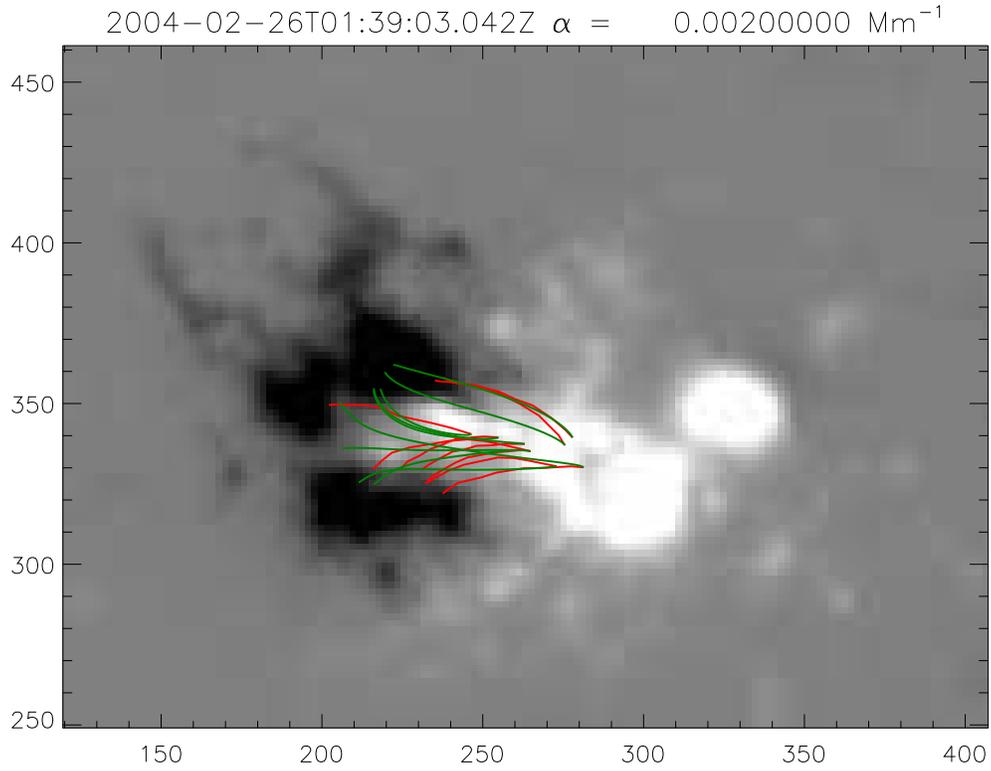}}
\caption{Observed coronal loops and approximations from a LFFF extrapolation.  The grey scale shows a section of the line-of-sight MDI magnetogram from before the flare (1:39).  Red curves show eight different loops traced from TRACE 171\AA\ images from at or around 2:13, rotated back to the time of the magnetogram.  Green curves are field lines from a LFFF extrapolated from the magnetogram.}
	\label{fig:loop_view}
\end{figure}

Some properties of the retracting flux tubes may be inferred from observations of the post-flare loops.  Figure \ref{fig:loop_view} shows several field lines (green) which approximate loops found in TRACE 171 \AA\ images from at or around 2:13 (see upper right panel of \fig\ \ref{fig:trace}).  We perform a linear force-free field (LFFF) extrapolation, from a preflare line-of-sight MDI magnetogram (grey scale).  We find that a twist parameter $\alpha=2\times10^{-11}\, {\rm cm^{-1}}$, yields field lines (green) most closely resembling the observed loops (red).  The resemblance is imperfect, but both form two families, and have roughly similar lengths.

\begin{figure}[htbp]
\centerline{\includegraphics[width=6.0in]{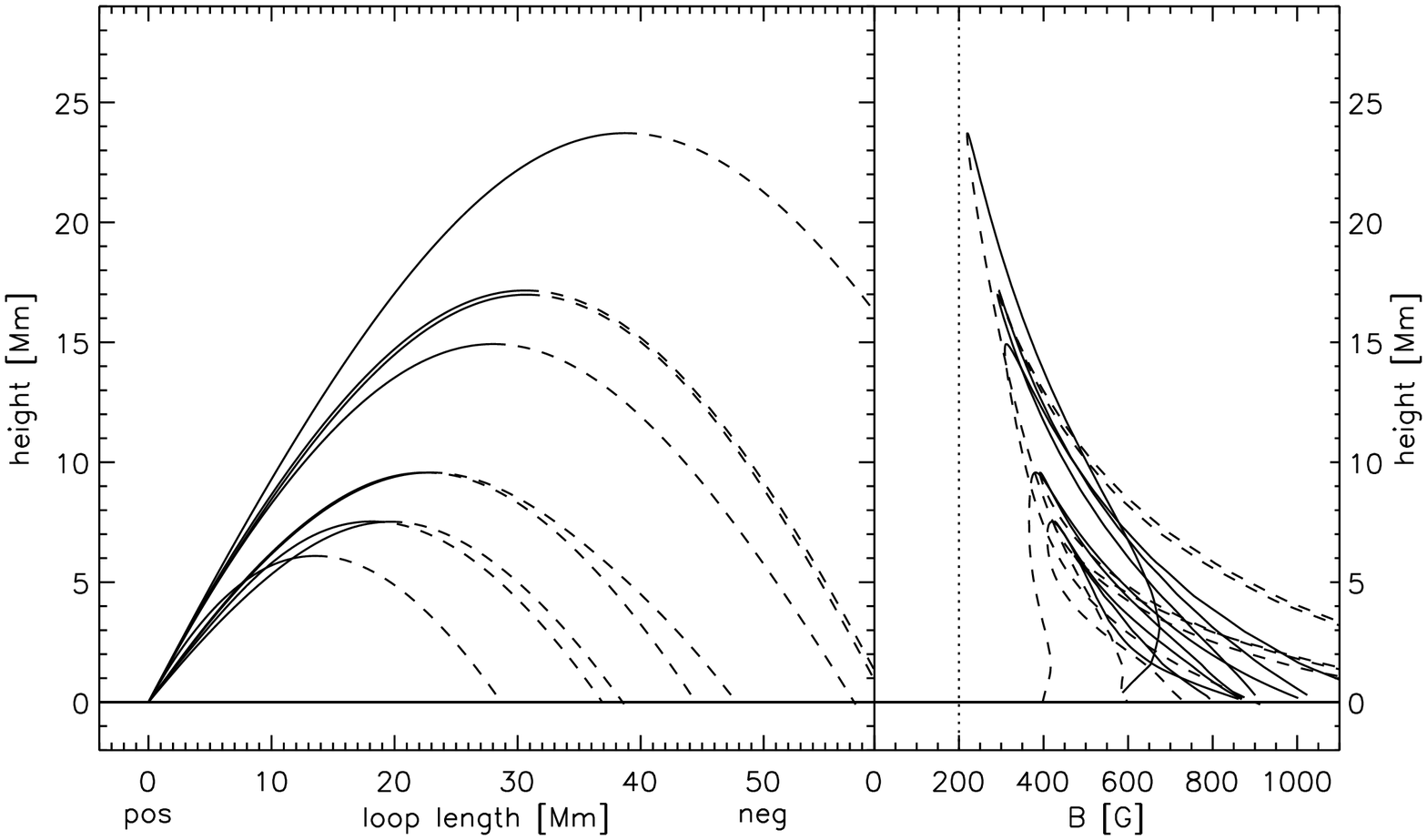}}
\caption{Properties of field lines from the LFFF plotted against the integrated length coordinate $\ell$.  The left panel shows the height in Mm, and the right panel shows the field strength in Gauss.  The positive (negative) leg of each loop is plotted using a solid (dashed) curve.}
	\label{fig:loop_prof}
\end{figure}

The observed loops had cooled to $\sim10^6$ K over $\sim10$--15 minutes after being formed by reconnection and retraction.   These loops would have been formed right at the peak of the flare, $\sim$2:00.  We therefore take the extrapolated field lines as the {\em final} state of our retracted flux tube.  Figure \ref{fig:loop_prof} shows the properties of the different field lines plotted in green in \fig\ \ref{fig:loop_view}.  Ideally we would perform a separate TFT simulation for each of the observed loops, but for this preliminary investigation we define a single loop with properties representative of the set.  We choose an intermediate loop length $L\simeq45$ Mm.  The field strength within the current sheet should be representative of the field {\em above} that loop.  We take this to be $B\simeq200$ G, which is a lower bound on the field strength of the post-flare loops.

\begin{figure}[htbp]
\centerline{\includegraphics[width=6.5in]{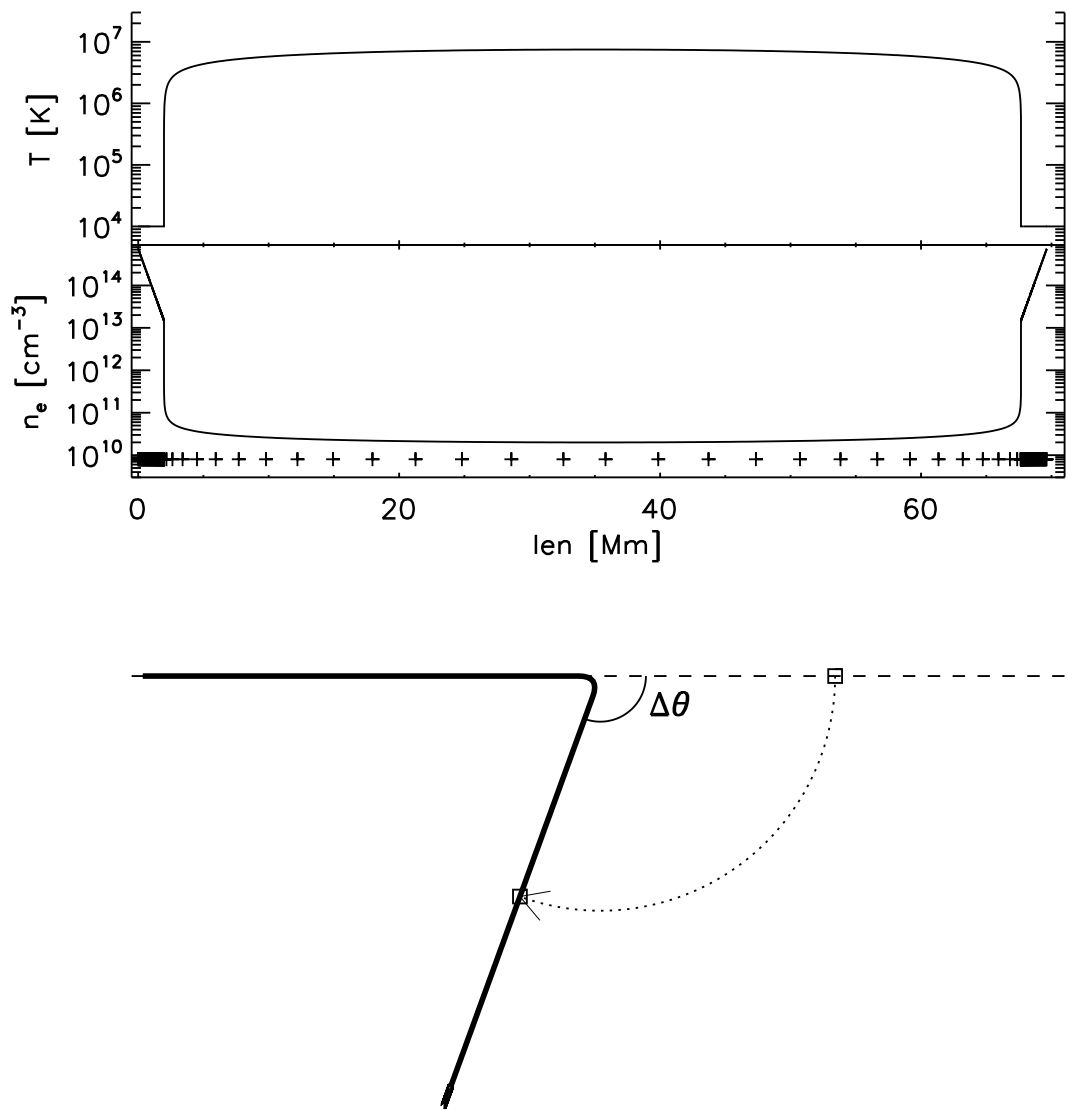}}
\caption{Plots of the temperature (top panel) and electron number density (middle panel) {\em vs.} arc length from one end for the initial loop.  Crosses below the density show the position of 
{\em every tenth} grid point.  The bottom panel shows how the tube is bent by an angle $\Delta\theta$ about its mid-point from its straight initial state (dashed) to a bent final state (thick solid).
The bent tube is then rotated counter-clockwise by $\Delta\theta/2$ (not shown) so that dynamical retraction will occur downward.}
	\label{fig:grid_demo}
\end{figure}

Based on the arguments above we set the final retracted length of our tube to be $L_{\rm fin}=45$ Mm.  The tube will have retracted through a current sheet whose strength should be representative of a super-arcade current sheet: $B=200$ G.  Having thus fixed the final state of the loop in our TFT model, we need to determine its {\em initial} state.  For convenience we begin with a flux tube in equilibrium --- i.e.\ an isobaric Rosner, Tucker and Vaiana (RTV) loop \citep{Rosner1978}, 
as shown in the upper panels of \fig\ \ref{fig:grid_demo}.  This equilibrium state is specified by its length, $L_0$, and apex temperature, $T_{0,{\rm max}}$.  The equilibrium loop is then bent by $\Delta\theta$, 
as indicated in the bottom panel of \fig\ \ref{fig:grid_demo}, with no effect on its plasma properties.  For simplicity it is bent about its mid-point 
(\citet{Longcope2015b} consider a more realistic case of bending about a off-center point, and find interesting but subtle differences).  
The bent tube is placed within the $x$--$z$ plane, which is taken as the plane of the CS.\footnote{Plots of the $x$--$z$ plane, such as the top of \fig\ \ref{fig:bend_plot} show a {\em side view} of the CS, rather that the more conventional {\em end-on} view.  The latter would appear as a very narrow line, or wedge, in the $y$--$z$ plane, and would prove far less informative.}  
It is positioned within the plane to make it symmetric about the vertical so that current flows horizontally ($\xhat$) and retraction occurs downward ($-\zhat$).

After performing runs at different values of these free parameters, we found that values $L_0=70$ Mm, $T_{0,{\rm max}}=7.5$ MK, and $\Delta\theta=110^{\circ}$ yielded reasonably good agreement with the amplitude and shape of the RHESSI HXR spectrum during the peak of the flare.  The details of this particular initial condition are given in table \ref{tab:IC}.  We return  in a later section to explain and motivate these particular parameters.

\begin{table}[htbp]
\centering
\begin{tabular}{llll}
\multicolumn{2}{l}{\sc Run parameters\dotfill} & \\ \hline
field strength & $B$ & 200 & G\\
initial full tube length & $L_0$ & $69.6$ & Mm \\
final full tube length & $L_{\rm fin}$ & $44.8$ & Mm \\
apex temperature & $T_{\rm max,0}$ & $7.5\times10^6$ & K \\
bend angle & $\Delta\theta$ & $110^{\circ}$ \\[9pt]
\multicolumn{2}{l}{\sc complete tube\dotfill} & \\ \hline
grid points & $N$ & 3556 \\
minium initial cell & $\delta\ell$ &78 & m \\[9pt]
\multicolumn{2}{l}{\sc initial corona\dotfill} & \\ \hline
coronal grid & $N_{\rm cor}$ & 297 \\
full length & $L_{0,{\rm cor}}$ & $65.7$ & Mm \\
pressure & $p$  & 34.8 & ${\rm erg\,cm^{-3}}$ \\
heating & $H_{\rm eq}$  & $9.2\times10^{-2}$ & ${\rm erg\, cm^{-3}\, s^{-1}}$ \\
$n_e$ at apex & & $1.7\times10^{10}$ & ${\rm cm}^{-3}$ \\
$v_a$ at apex  & & 3,090  & ${\rm km/s}$ \\
\multicolumn{2}{l}{electron column of half-loop}  & $7.2\times10^{19}$ & ${\rm cm^{-2}}$ \\[9pt]
\multicolumn{2}{l}{\sc initial chromosphere\dotfill} & \\ \hline
$T_{\rm min,0}$ & & $1.0\times10^4$ & K \\
depth & & 2.0 & Mm \\
$n_e$ at top & & $1.3\times10^{13}$ & ${\rm cm}^{-3}$ \\
$n_e$ at base & & $6.4\times10^{14}$ & ${\rm cm}^{-3}$ \\
\multicolumn{2}{l}{total electron column}  & $3.2\times10^{22}$ & ${\rm cm^{-2}}$ \\
\multicolumn{2}{l}{number of scale heights} & 3.9 \\
maximum $\delta\ell$ & & 44.8 & km 
\end{tabular}
\caption{Properties of the numerical simulation.}
	\label{tab:IC}
\end{table}

The tube configuration described above is used to initialize a flux tube axis modeled with $N=3,556$ Lagrangian grid points.  The grid points are distributed to keep roughly similar mass between every pair 
of points,
as indicated by $+$s in the middle panel of \fig\ \ref{fig:grid_demo}. 
As a result only 297 points are initially in the corona, while the remainder form a stratified $10^4$ K chromospheres, 2.0 Mm deep, at each of the tube's feet.  (\citet{Longcope2015b} provide more details about the construction of this chromospheric layer.)

\subsection{The TFT equations}

The dynamical evolution of the retracting flux tube is solved using the PREFT code described in \citet{Longcope2015b}.  Its Lagrangian grid points describe the tube's axis, $\rvec(\ell,t)$, where the length from one end, $\ell$, must be recomputed at each instant.  Each Lagrangian point moves at the plasma velocity $\vvec(\ell,t)$, which itself changes according to the momentum equation
\begin{eqnarray}
  \rho{d\vvec\over dt} &=& \left( {B^2\over 4\pi} - p \right)\, {\partial\lhat\over\partial\ell} 
  ~-~ \lhat\, {\partial p\over \partial \ell} 
   ~+~ \hbox{${4\over3}$}\,{\partial\over\partial\ell}\left[ \, \lhat\,\mu\, \left(\lhat\cdot
  {\partial \vvec\over \partial\ell}\right)\, \right] ~+~ \rho\,\gvec ~~,
  	\label{eq:mom}
\end{eqnarray}
where $\lhat=\partial\rvec/\partial\ell$ is the tangent vector to the axis, $\rho$, and $p$  are the mass density, and pressure, $\gvec$ is the gravitational acceleration, and $\mu$ is the dynamical viscosity.  The first term on the right hand side (rhs) is the magnetic tension term, from the Lorentz force, and is responsible for shortening the tube, and thereby converting magnetic energy into bulk plasma motion.  The second and third terms, pressure and viscosity, are responsible for developing and resolving the SMSs at which some of the bulk kinetic energy is thermalized.

Instead of advancing a continuity equation, the Lagrangian code uses mass conservation to compute mass  density directly.  Between a pair of adjacent grid points lies a cell with a fixed mass per magnetic flux $\delta m$.  If the grid points are separated by $\delta\ell$ at some instant, the mass density in 
the cell is
\be
  \rho(\ell,t) ~=~ {\delta m\over \delta\ell}\, B ~~,
\ee
at that instant,  where $B$ is the fixed value of field strength imposed from outside the current sheet. 

The temperature $T$ within each cell is advanced using the energy equation \citep{Longcope2015b}
\begin{eqnarray}
    \cv \rho \, {dT\over dt} &=&
  - p\,\lhat\cdot{\partial\vvec\over\partial\ell} 
  ~+~ {\partial\over\partial\ell}\left( \, \kappa\, {\partial T\over \partial\ell} \right)
 ~-~ n_e^2\Lambda(T) 
  ~+~ \hbox{${4\over3}$}\mu \left(\lhat\cdot{\partial \vvec\over \partial\ell}\right)^2 ~+~h ~~,
  	\label{eq:erg}
\end{eqnarray}
where $\cv$ is the specific heat, $\kappa$ is the thermal conductivity, $n_e$ is the electron number density, and 
$\Lambda(T)$ is an optically thin radiative loss function.  The penultimate term on the rhs is the heating resulting from the viscous dissipation in the momentum eq.\ (\ref{eq:mom}).  This is the source of flare heating in the model, so every effect we predict ultimately originates in the conversion of magnetic energy.  A very small {\em ad hoc} heating, $h$, is required to maintain the loop in its initial RTV equilibrium, but plays no noticeable role in the flare.

We assume a fully ionized plasma, so the mean particle mass is
$\bar{m} = 0.593\, m_p$, the specific heat is $\cv=(3/2)\kb/\bar{m}$, 
and the electron number density is $n_e=0.874(\rho/m_p)$, where,  $m_p$ is the proton mass, and 
$\kb$ is Boltzmann's constant.  Pressure in the cell is found from the ideal gas law, $p=(\kb/\bar{m})\rho T$.

For modest temperature gradients thermal conductivity would be given by the classical Spitzer-H\"arm version
\be
  \kappa_{\rm sp}(T) ~=~ \kappa_0\, T^{5/2} ~~,
  	\label{eq:kappa_sp}
\ee
with $\kappa_0=10^{-6}$ in cgs units.   When temperature gradients are very large the classical heat flux, 
$\kappa_{sp}|\partial T/\partial\ell|$, would exceed the free-streaming flux 
\be
  F_{\rm fs} ~=~ \hbox{${3\over2}$}\,n_e\,\kb T\,v_{\rm th,e} ~~,
\ee
where $v_{\rm th,e}=\sqrt{\kb T/m_e}$ is the electron thermal speed ($m_e$ being the electron mass).  We keep the heat flux below some fraction of this theoretical free-streaming limit, $\xi F_{\rm fs}$, by using the modified conductivity
\be
  \kappa ~=~ \kappa_{\rm sp}\, \left[\, 1 ~+~ \left( {\kappa_{\rm sp}|\partial T/\partial\ell|\over \xi\, F_{\rm fs}}
  \,\right)^2\,\right]^{-1/2} ~~,
    	\label{eq:kappa}
\ee
in the energy equation (\ref{eq:erg}).  Following \citet{Longcope2014b}, and \citet{Longcope2015b},  we adopt the value $\xi=1$.  The viscosity is always computed from the classical form
\be
  \mu(T) ~=~ {\rm Pr}\, {\kappa_{\rm sp}(T)\over \cv} ~~,
\ee
where ${\rm Pr}\simeq 0.012$ is the Prandtl number of a fully ionized plasma.

While the TFT differs in many respects from the classic reconnection model of Petschek \citep{Petschek1964,Vasyliunas1975,Soward1982b,Skender2003}, it shares some of its basic features.  In particular, energy conversion in both occurs {\em after} the reconnection, as field line retraction creates shocks which heat and compress the plasma.   As a result TFT and the Petschek model predict similar temperatures and densities in the reconnection outflow \citep{paperI,Longcope2011b,Longcope2015b}.

\subsection{Post-retraction evolution}

\begin{figure}
\includegraphics[width=7.6in]{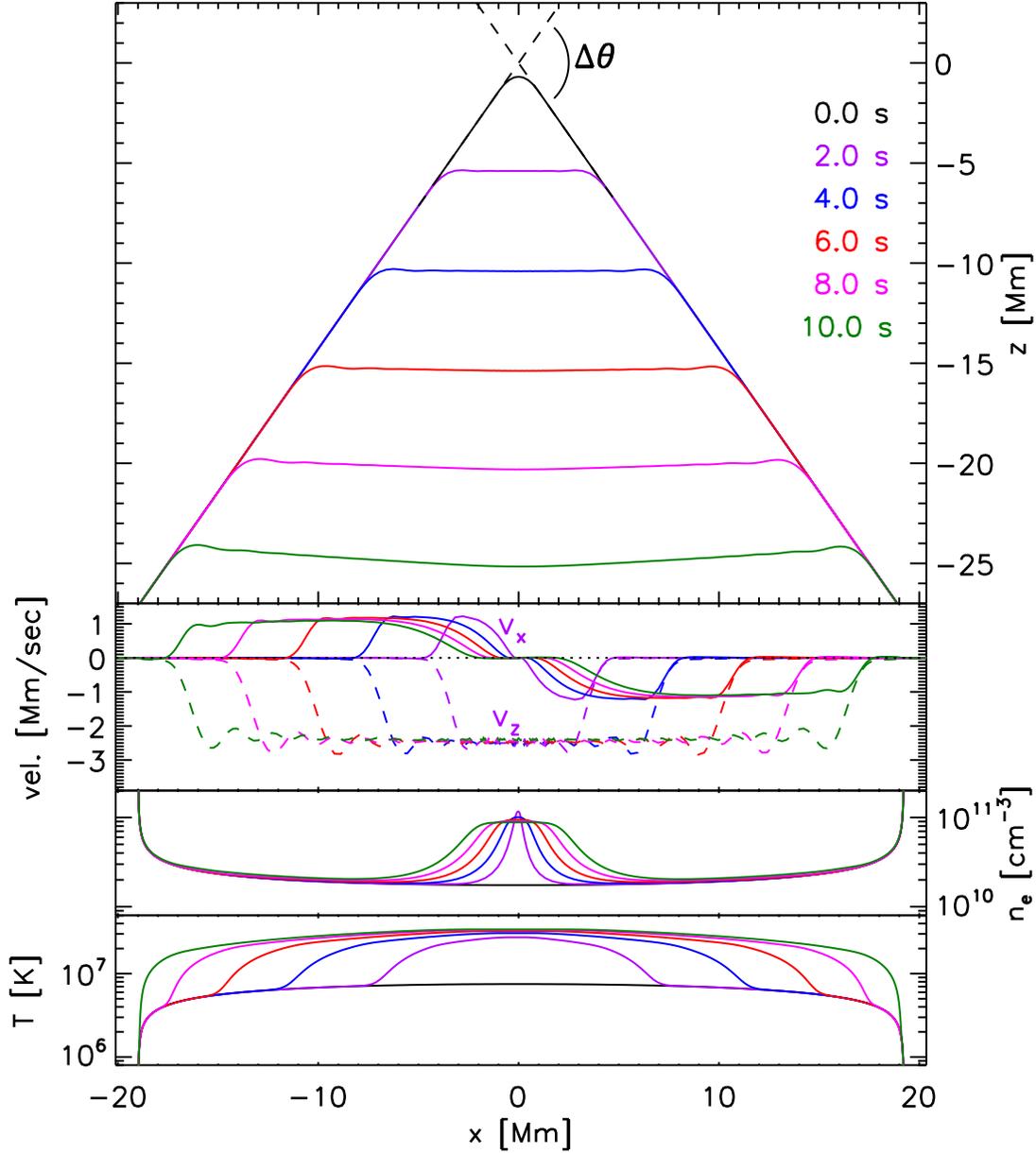}
\caption{Post-reconnection retraction of the flux tube. Top panel shows the axis of the flux tube in a side view of the planar CS.  Dashed black lines show the straight tubes whose reconnection produced the bent initial flux tube (solid black).  Times are coded by color.  Below this are plotted the horizontal (solid) and vertical (dashed) plasma velocity {\em vs.} horizontal coordinate $x$.  The bottom two panels show electron density and temperature (bottom) on logarithmic scales.}
	\label{fig:bend_plot}
\end{figure}

The post-retraction evolution follows from solving these dynamical equations beginning with the bent flux tube described above, 
{shown in black in the top panel of \fig\ \ref{fig:bend_plot}, and matching the configuration shown in the bottom panel of \fig\ \ref{fig:grid_demo}.  
This represents the configuration just after the reconnection of two straight tubes (dashed lines) from opposite sides of the CS.}
{The subsequent evolution is shown in colors in \fig\ \ref{fig:bend_plot},}
resembles those reported in previous TFT simulations \citep{Guidoni2010,Guidoni2011,Longcope2011b,Longcope2015b}.  
Rotational discontinuities (RDs) appear as bends in the axis propagating at the Alfv\'en speed, $v_a\simeq3.1$ Mm/s.  These produce a central horizontal segment moving downward with vertical velocity 
$v_z=-v_a\sin(\Delta\theta/2)\simeq-2.5$ Mm/s: the retraction.  They also produce horizontal flows within the segments.  The parallel velocity from a pure RD would be, 
$v_x=\pm2v_a\sin^2(\Delta\theta/4)\simeq\pm 1.3$ Mm/s, which is close to the value actually achieved.  These transient, field-aligned flows are directed inward, toward the loop's center, and are here called {\em compression flows}; they are not equivalent to the ``reconnection inflows'' of steady models, which would be directed into or out of the page in \fig\ \ref{fig:bend_plot}.

\begin{figure}
\includegraphics[width=6.5in]{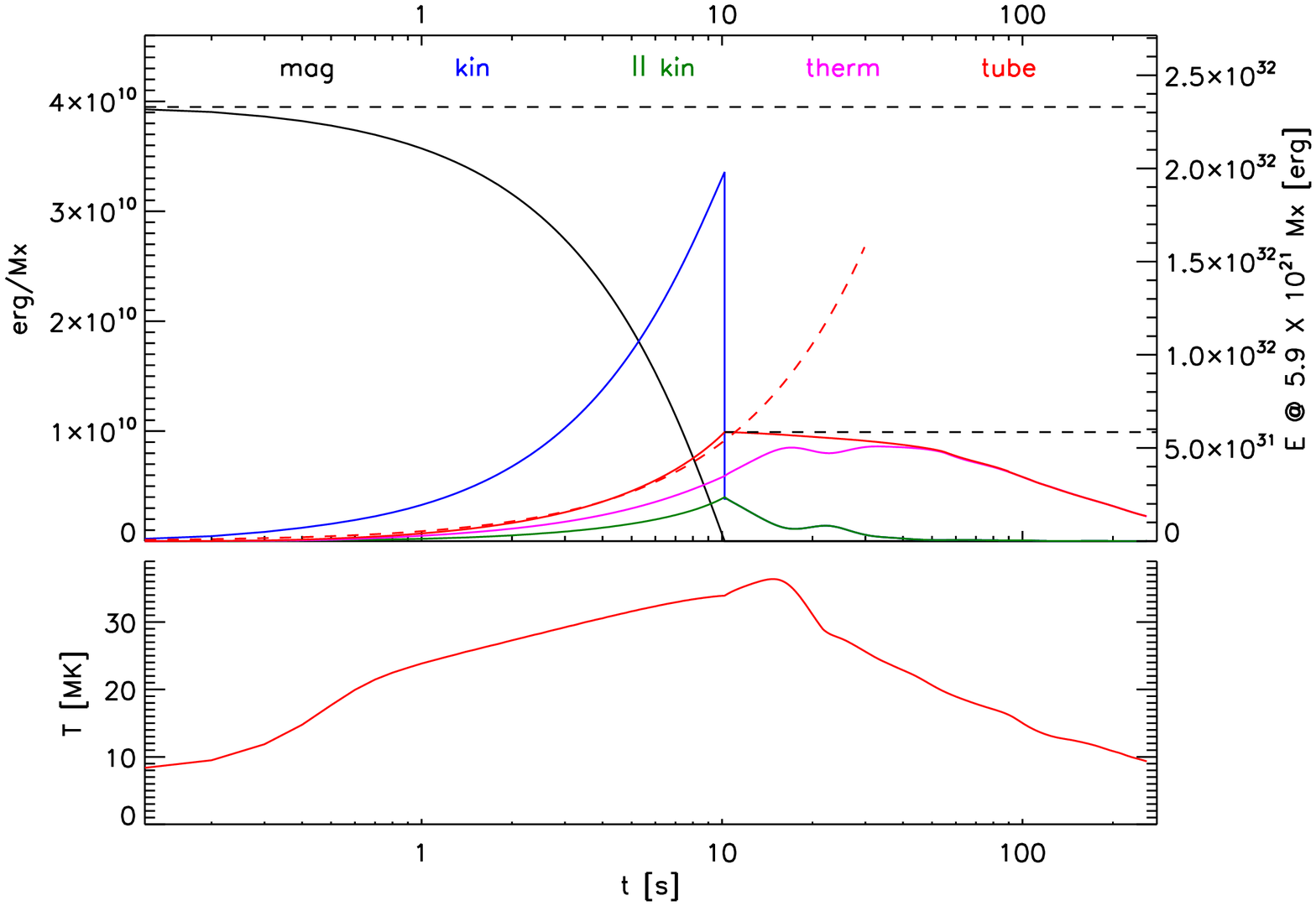}
\caption{Summary of the time evolution of the TFT simulation.  The top panel shows different energies in units of erg/Mx (left axis) {\em vs.} a logarithmic time axis.  This is converted, on the right axis, to a total energy by multiplying by the total reconnected flux $\Delta\Phi_{\rm rib}=5.9\times10^{21}$ Mx.  Plotted are the free magnetic energy (black), total kinetic energy (blue), kinetic energy in purely parallel flows (green), and thermal energy energy above initial value (magenta).  The thermal and parallel kinetic energies are confined to the tube and their sum is plotted as the {\em tube} energy (red).  Peak values of total and tube energies are indicated by dashed horizontal lines.  The bottom panel plots peak temperature in MK against the same time axis.}
	\label{fig:erg_plot}
\end{figure}

The shortening of the flux tube releases magnetic energy (per unit flux) at a rate
\be
  \dot{W}_M ~=~ {B\over 4\pi}{dL\over dt} ~=~ {B\over 4\pi}\, 4v_a\,\sin^2(\Delta\theta/4)
  ~\simeq~ 4.2\times10^{9}\, {\rm erg/s/Mx} ~~,
  	\label{eq:WMdot}
\ee
as shown by the black curve in \fig\ \ref{fig:erg_plot}.
The RDs convert this released energy into kinetic energy (blue) of both vertical and horizontal motions.  The kinetic energy of horizontal motion (green) accounts for a fraction, 
$v_x^2/v^2=\sin^2(\Delta\theta/4)\simeq20\%$, of the energy released.  This fraction is thermalized at the slow magnetosonic shocks visible as a density peak in the third panel of \fig\ \ref{fig:bend_plot}.  The shocks thus produce a flare energy flux
\be
  F_{\rm fl} ~\simeq~ \half B\,\dot{W}_M\, \sin^2(\Delta\theta/4) ~=~
  0.9\times10^{11}\, {\rm erg/cm^2/s} ~~,
\ee
into each leg of the flare loop.  This is the energy of the flare.

The thermalized energy moves away from the shocks conductively, since non-thermal electrons are not part of the model.  This produces a conduction front which, had the flux limiter been dominant, would propagate at \citep{Longcope2015b}
\be
  v_{\rm fr} ~\simeq~ \xi\, f_e\, \left({2\over 3}\, {F_{\rm fl}\over m_e n_e}\,\right)^{1/3}
  ~\simeq~ 8.2\, {\rm Mm/s} ~~,
  	\label{eq:vfr}
\ee
where $f_e=0.52$ is the fraction particles which are electrons.  Since this exceeds the Alfv\'en speed, $v_a\simeq3.1$ Mm/s, the conduction fronts move ahead of the RDs, as is evident in \fig\ \ref{fig:bend_plot}.  At this speed they would reach the chromosphere in roughly 
$L_{\rm cor}/2v_{\rm fr}\simeq4.3$ s.  
{It appears, however, that the flux limiter does not play a dominant role, as we had assumed.   The actual heat flux reaches a maximum of $0.43F_{\rm fs}$ at $t=0.7$.  By $t=3$ sec points 9 Mm to either side of the apex have fallen to $0.25F_{\rm fs}$, are thereafter the peak flux drops below that value.  As a consequence the front propagates at about half the estimate in \eq\ (\ref{eq:vfr}) and reaches the chromosphere at about $t=10$ sec.}

The flare energy flux, $F_{\rm fl}$ seeks to drive the loop apex to a final temperature \citep{Longcope2014b}
\be
  T_{\rm fl} ~\simeq~ 1.46\left({F_{\rm fl}L_{\rm cor}/2\over\kappa_{\rm 0,sp}}\right)^{2/7}
  ~\simeq~ 4.8\times10^7\, {\rm K} ~~,
\ee
using the final length of the coronal loop segment, $L_{\rm cor}=45$ Mm.  The apex temperature has risen to $T\simeq3\times10^7$ K by the time the conduction fronts reach the chromosphere, but after that the retraction is halted and the asymptotic flare temperature is never achieved (see the red curve in the bottom panel of \fig\ \ref{fig:erg_plot}).

At $t=10.2$ s the loop has shortened from it initial total length $L_0=69.6$ Mm, 
to $L_{\rm fin}=45$ Mm, close to the target value from the field-line extrapolation.  The retraction is then halted by straightening the tube and zeroing the velocity components perpendicular to the axis.
{This is intended to simulate the arrival of the retracting tube at the top of the post-flare arcade, at which point retraction ceases.  Such a cessation was achieved more realistically by \citet{Guidoni2011}, but a later investigation by \citet{Longcope2015b} showed that an abrupt {\em ad hoc} straightening can produce similar results.  This method is used here.  The perpendicular velocity of the tube is immediately stopped, while all parallel motion persists.}

\begin{figure}
\centerline{\includegraphics[width=6.8in]{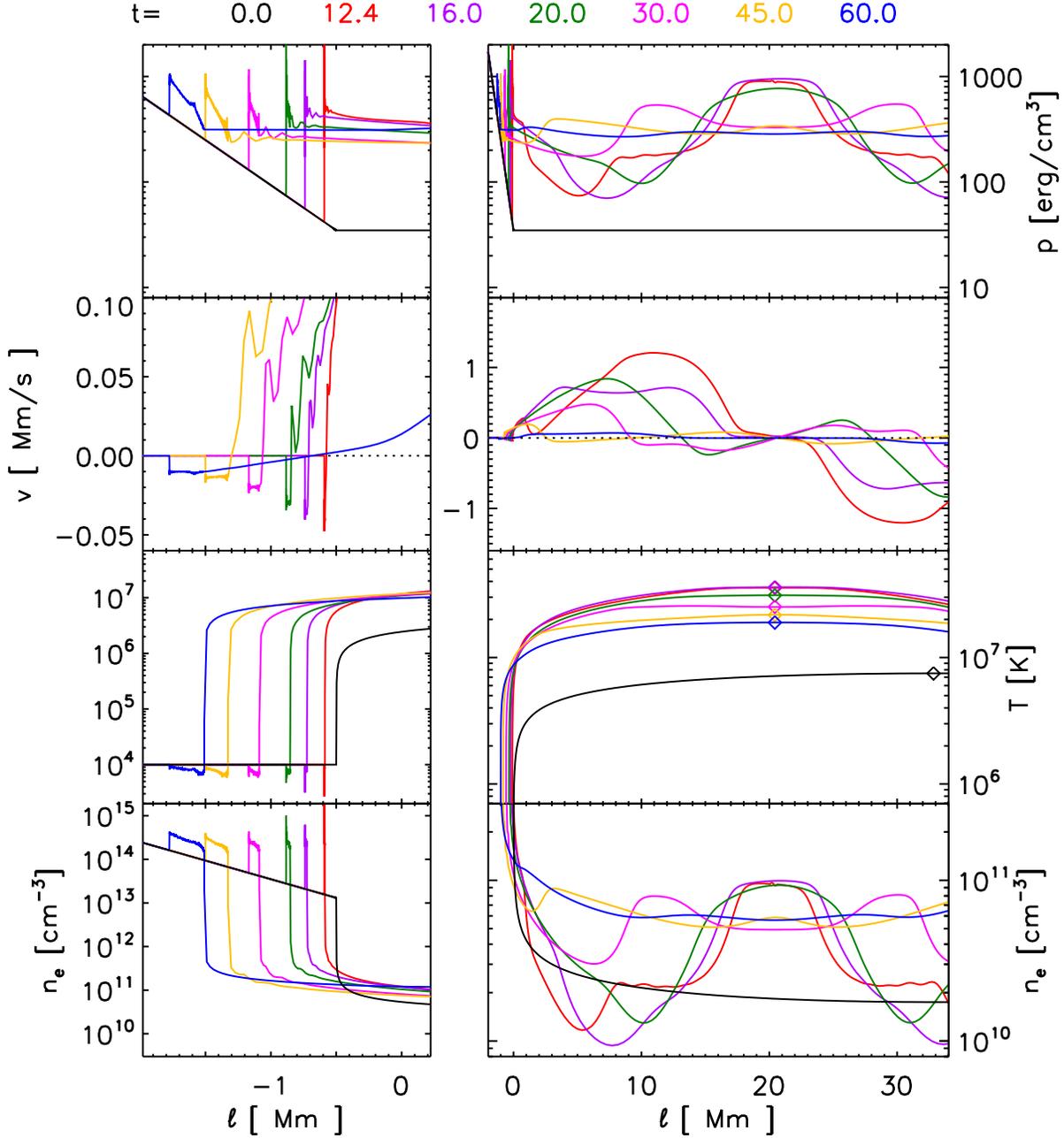}}
\caption{Evolution after the flux tube is straightened.  Rows from top to bottom show pressure, 
velocity, temperature, and electron density {\em vs.} $\ell$ measured from the initial transition region.  Plots are made from a selection of different times using colors defined along the top.  The right column shows the entire left section of the tube, while the left column shows an expanded view of the chromosphere.  The black curve corresponds to the initial condition, $t=0$, when the tube had its full length, $L_0=70$ Mm, while in all others it has achieved its final length $L_{\rm fin}=45$ Mm.  Diamonds on the $T(\ell)$ plots show the tube's mid-point, which shifts leftward due to retraction.}
	\label{fig:evap_plot}
\end{figure}

The evolution past this time, shown in \fig\ \ref{fig:evap_plot}, occurs in a straight tube whose parallel velocity was initiated by the retraction phase.  This persistent parallel velocity is able to maintain the slow magnetosonic shocks, i.e. the loop-top density concentration, despite the absence of RDs.  The bottom panel of \fig\ \ref{fig:evap_plot} shows this concentration of 
$n_e\simeq10^{11}\,{\rm cm^{-3}}$ at $t=16$ s being maintained by diminished ($v_x\simeq\pm600$ km/s) compression flows.  The central concentration eventually overcomes the persistent compression flow to disassemble itself.  By $t=20$ the flows have reversed to $v_x\simeq\mp300$ km/s, although the peak density has not yet shown signs of reducing.    Thus the persistent remnants of the SMSs rest atop the now static loop, as originally predicted by \paperI, for at least 10 seconds beyond the end of retraction.

At the same time the compression flows are maintaining the loop-top concentration ($t=16$ s), evaporation fronts have propagated $\ell\simeq 3$ Mm from the chromosphere. The evaporation flow forms a second peak, $v_x\simeq\pm800$ km/s, behind that from the retraction-generated compression flow.  Moving at that speed, the opposing evaporation fronts would reach the apex ($\ell=L_{\rm fin}/2=22.5$ Mm) and collide at $t\simeq28$ s.    By $t=30$ s the evaporation fronts have collided with the central density concentration and been reflected backward.  The result is a pair of peaks $n_e=7\times10^{10}\,{\rm cm^{-3}}$ located at $\ell\simeq10$ and $32$ Mm, and moving outward.  By $t=45$ s these reflected peaks have almost reached the chromosphere once more (the left peak is at $\ell=3$ Mm, while the right has moved off the plot).

After the foregoing collision and reflection the plasma motion substantially diminishes.  By $t=60$ s, the parallel flows have largely subsided and the loop has achieved an approximately isobaric equilibrium with very little flow.  The tube continues cooling by radiation, remaining close to an isobaric equilibrium, until by $t=260$ s its peak temperature has fallen to $10$ MK (see bottom panel of \fig\ \ref{fig:erg_plot}), and the run is ended.

The transition from retraction to rest is evident in the energy plot of \fig\ \ref{fig:erg_plot}.  At $t=10.2$~s the tube has fully retracted, it is straightened, and perpendicular flows are artificially halted, causing the kinetic energy to drop.  The parallel flow, whose energy is plotted in green, is unaffected by the straightening, and thereafter constitutes all of the kinetic energy.  The parallel kinetic energy and thermal energy (magenta) are confined to the tube, and compose the total flare energy, plotted in red.  This energy rises approximately linearly at the rate $\dot{W}_M\sin^2(\Delta\theta/4)$, plotted as a red dashed curve.  The complete retraction therefore introduces 
$\Delta W_M\sin^2(\Delta\theta/4)=8.4\times10^{9}$ erg/Mx into the flux tube, partitioned into comparable parts thermal and kinetic energy.  The residual kinetic energy is converted to thermal energy in the course of maintaining the apex concentration.  This conversion halts at $t\simeq17$~s when the compression flows reverse and the apex temperature peaks 
at $T=36$~MK.  Thereafter the thermal energy stored in the loop top is converted to kinetic energy of expansion, and then back to thermal upon reflection at the chromosphere.   By $t=40$ these oscillations have largely ceased and the loop begins cooling quasi-statically through radiative losses.

\subsection{The super-hot source}

\begin{figure}[htbp]
\centerline{\includegraphics[width=7.5in]{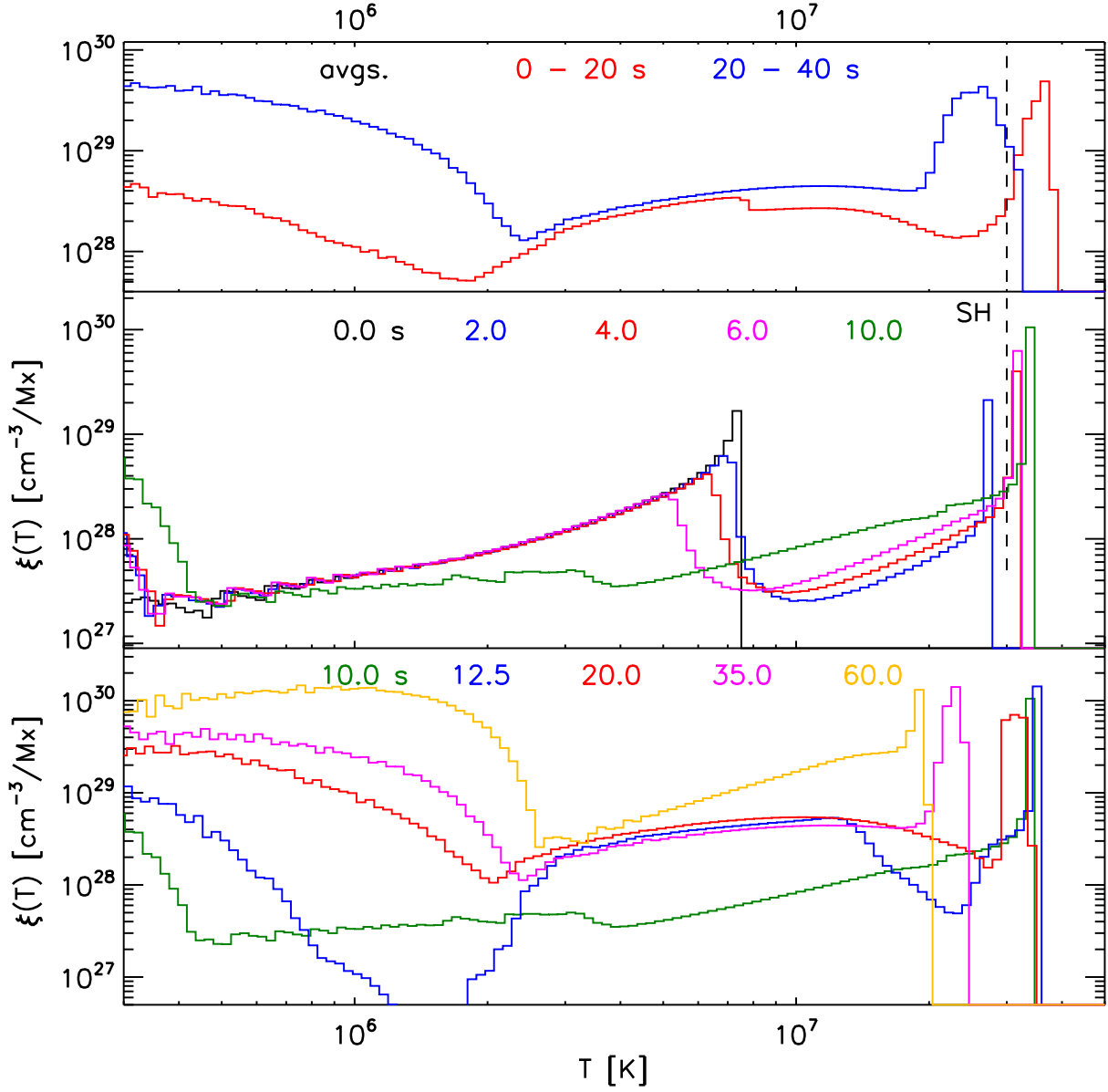}}
\caption{The evolution of the logarithmic DEM, $\xi(T)$, defined in \eq\ (\ref{eq:DEM}).  The middle panel shows early times during the retraction using colors defined along the top.  A vertical dashed line delineates the portions above $T=30$ MK, designated super-hot (SH).  The lower panel gives times later in, and following, the retraction.  The top panel shows the DEM averaged over two 20~s intervals: 0--20 s (red) and 20--40s (blue).}
	\label{fig:DEM}
\end{figure}

Observable consequences of the evolution described above, including a super-hot source, can be anticipated from the evolution of the differential emission measure (DEM).  To investigate this we compute a logarithmic DEM per unit flux
\be
  \xi(T_j,t) ~=~ \int\, n_e^2(\ell,t)\, S_j\bigl[\, T(\ell,t)\,\bigr]\, {d\ell\over B} ~~.
  	\label{eq:DEM}
\ee
The logarithmic DEM at temperature $T_j$ centering a bin of width $\Delta T_j$ is produced using the indicator function
\be
  S_j(T) ~=~ \left\{\begin{array}{lcl} \ln(10)\displaystyle{T_j\over \Delta T_j} &~~,~~&
  |T-T_j|<\half\Delta T_j \\ 0 &~~,~~& \hbox{otherwise} ~~.\end{array} \right.
\ee
The integration in \eq\ (\ref{eq:DEM}), uses differential volume per flux, $d\ell/B$, giving $\xi(T,t)$ units of 
${\rm cm}^{-3}$ per Maxwell.  It is plotted in \fig\ \ref{fig:DEM} for a selections of times corresponding to samples from \figs\ \ref{fig:bend_plot} and \ref{fig:evap_plot}.

A super-hot (SH) source develops during the retraction phase, shown in the middle panel of \fig\ \ref{fig:DEM}.  The slow shocks create a narrow DEM peak around $T=32$~MK by compressing and heating  the loop-top coronal material, thus creating a DEM peak while at the same time eroding away the right end of the preflare DEM, shown in black.  The DEM below this value steadily rises as the conduction fronts move into lower, and therefore denser, portions of the loop.  This phase ends when those conduction fronts reach the chromosphere which, as noted in \fig\ \ref{fig:bend_plot}, has begun by $t=10$~s.  In the corresponding DEM (green) the emission below 0.5 MK has increased as material at chromospheric density is heated to coronal temperatures.  By $t=12.5$~s the evaporative up flows have raised the DEM over the region $3<T<12$ MK.

The narrow DEM peak created by the SMSs is separated, by a DEM trough, from the broad evaporation flow peak.  This separation persists until $t\simeq20$~s (red curve in lower panel).  This is the time at which the evaporation flows have nearly reached the central density concentration (see \fig\ \ref{fig:evap_plot}).  By $t=35$ s that collision has occurred, and the SH peak has merged with the evaporation, reducing the peak to a more modest temperature of $T\simeq 22$~MK.  

Even if this rapid time evolution is not observationally resolvable, the SH source is still evident.  The top panel of \fig\ \ref{fig:DEM} shows the DEM averaged over 20 s intervals.  The first average (0--20 s, in red) shows a distinct but broadened SH peak to the right of the dashed $t=30$ MK line.  In the second average (20--40 s, in blue) the peak has moved leftward owing to the cooling effects of evaporation; the source is no longer super-hot.

The DEM can be used to estimate the emission measure of the SH source.
The peak's peak value is $\xi\simeq10^{30}\,{\rm cm^{-3}/Mx}$, from $t=10$ to 20~s.  The integrated emission measure of this very narrow peak, above 30 MK (dashed vertical line), is 
$\simeq2\times10^{28}\,{\rm cm^{-3}/Mx}$, by $t=10$ s, and rises to $\simeq4\times10^{28}\,{\rm cm^{-3}/Mx}$, by $t=20$~s when the peak has broadened due to interaction with evaporation.  By $t=21$ s, however, evaporation has driven the peak temperature below 30 MK. Time integrating the integrated EM yields an emission measure per reconnection rate
\[
  \int dt \int d\,(\log T)\, \xi(T,t) ~\simeq~ 4.8\times 10^{29}\, {\rm {cm^{-3}\over Mx/s}} ~~,
\]
for material $T>30$ MK.
Multiplying by the instantaneous reconnection rate, $\dot{\Phi}_{\rm rib}$, yields the observed emission measure of plasma above $30$~MK.  The peak observed reconnection rate, 
$\dot{\Phi}_{\rm rib}\simeq1.6\times10^{19}$ Mx/s (see top of \fig\ \ref{fig:thist}), yields an emission measure, $EM=7\times10^{48}\,{\rm cm^{-3}}$, matching well the estimate of the RHESSI super-hot source given in \paperI.

\section{Synthesizing flare light curves}

The complete PREFT simulation provides the dynamical evolution of a single tube of flux following its reconnection.  It may be used to synthesize that single tube's contribution to a particular observable band of emission.  Characterizing that band by a temperature response per emission measure, denoted $R(T)$, the single-tube contribution is
\be
  I_0(t) ~=~ \int\, n_e^2(\ell,t)\, R\bigl[\, T(\ell,t)\, \bigr]\, {d\ell\over B} ~~.
  	\label{eq:I0}
\ee
As in \eq\ (\ref{eq:DEM}), this is an integral over volume per magnetic flux, $d\ell/B$, and thus results in a light-curve {\em per magnetic flux}.  If $n_e^2RV$ produced a count rate in photons/s, then $I_0(t)$ will have units photons/s/Mx.

The loop retraction modeled by PREFT resulted from a single unit of magnetic flux reconnecting at the instant $t=0$.  This means the light curve $I_0(t)$ is the response to an impulsive flux transfer with a transfer rate $\dot{\Phi}(t)=\delta(t)$, the Dirac-delta.  If flux were actually reconnected at some time-varying rate $\dot{\Phi}(t)$, then the total observed light curve would be produced by convolving it with the impulse response
\be
  I(t) ~=~ \int^t\, I_0(t-s)\, \dot{\Phi}(s)\, ds ~~.
  	\label{eq:conv}
\ee

\begin{figure}[htbp]
\centerline{\includegraphics[width=6.9in]{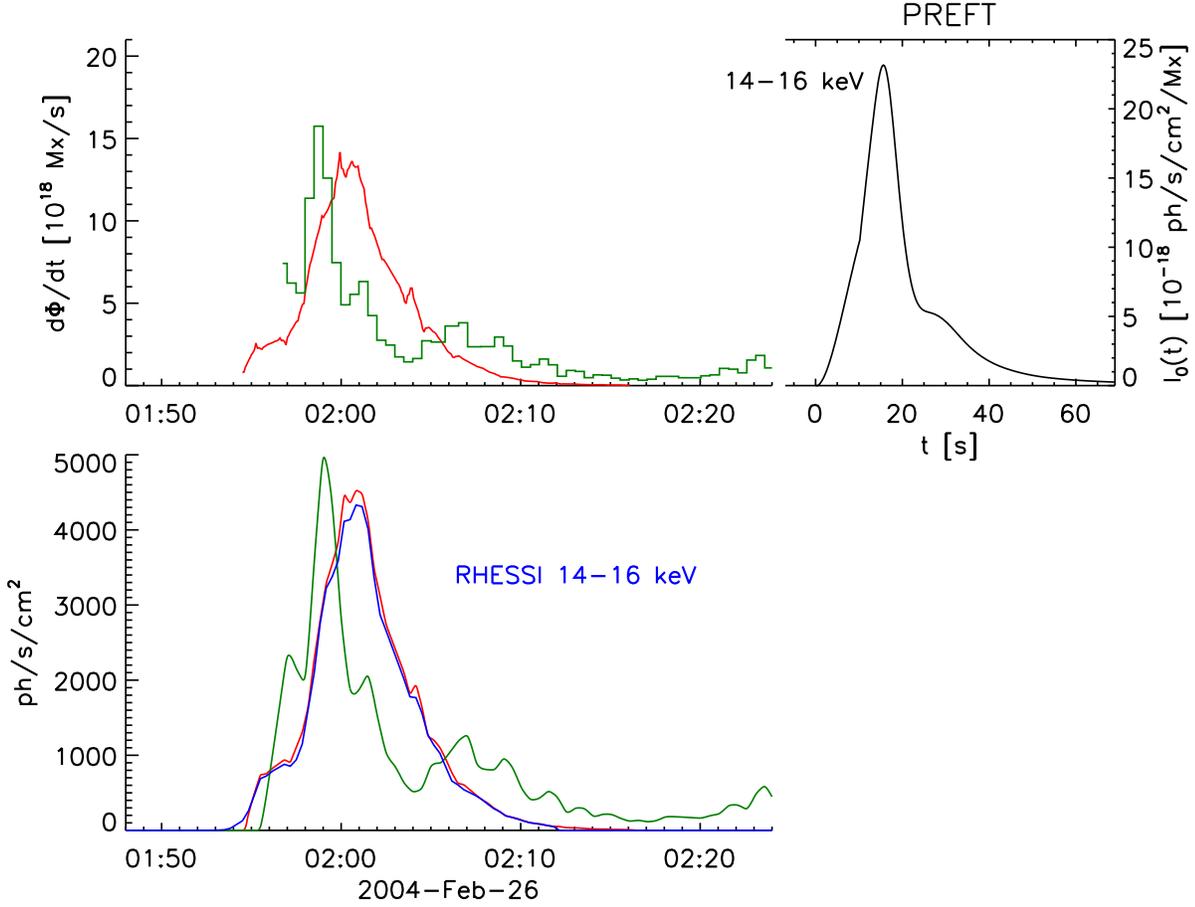}}
\caption{The synthesis of a light curve from an impulse response and flux transfer rate.  The impulse response, $I_0(t)$ (black curve, upper right), is derived from the PREFT run using \eq\ (\ref{eq:I0}) for a band of thermal brems{\-}strah{\-}lung over photon energies $14\, {\rm keV}<\varepsilon_{\gamma}<16\, {\rm keV}$ .  Flux transfer rates $\dot{\Phi}(t)$ (upper left) are found from ribbon motion (green) and inversion (red).  The convolution of each with $I_0(t)$ results in the synthetic light curves (green and red, respectively) shown in the lower left.  The blue curve is the curve derived from RHESSI data, and used to derive $\Phi_{\rm inv}(t)$ (red).}
	\label{fig:synth}
\end{figure}

To illustrate this synthesis we produce a response function $R(t)$ by integrating thermal brems{\-}strah{\-}lung emission \citep{Rybicki1979} over a narrow range of photon energies $14\, {\rm keV}<\varepsilon_{\gamma}<16\, {\rm keV}$ and evaluating the photon flux at 1 AU.  The product $n_e^2RV$ has units of ${\rm photons/s/cm^2}$.  Using this, and evolution of the above PREFT simulation, in eq.\ (\ref{eq:I0}) yields the impulse response $I_0(t)$ plotted in the upper right panel of \fig\ \ref{fig:synth}. This consists of a peak at $t=15$ s from the loop-top source produced by the slow mode shocks, followed by a more gradual decay after $t=25$~s, produced by the effects of evaporation.  The emission in this energy band has largely subsided by $t=60$~s, due to plasma cooling.

The flux transfer rate measured from ribbon motion, $\Phi_{\rm rib}(t)$, is plotted in green in the upper left panel of \fig\ \ref{fig:synth}.  The convolution of this with the impulse response produces a synthetic light curve $I(t)$ plotted as a green curve in the lower left panel.  This compares favorably, but not perfectly, with the photon flux measured by RHESSI, plotted in blue.  Both peak at similar levels $I\simeq 4500\, {\rm ph/s/cm^2}$, although the synthetic curve peaks about 2 minutes earlier than the RHESSI curve.  The synthetic light curve (green) shows a low-level tail beyond 2:10, when virtually all observed HXR emission has ceased.  These discrepancies are likely the result of imperfect measurement of the ribbon motion using 171 \AA\ data instead of a purely chromospheric line.  The later tail suggests that the loop properties vary over the flare, while our synthesis uses identical loops throughout.

An alternative to using the measured flare-ribbon flux is to invert the convolution operation in \eq\ (\ref{eq:conv}).  Using the observed RHESSI light curve, $I(t)$, and the impulse response $I_0(t)$, yields a flux transfer rate 
$\dot{\Phi}_{\rm inv}(t)$, plotted in red in the upper left panel of \fig\ \ref{fig:synth}.  This peaks at a level similar to 
$\dot{\Phi}_{\rm rib}(t)$, but slightly later.  Remarkably, the integrals of both curves are very similar: 
$\Delta\Phi=6\times10^{21}$ Mx.  This can be taken as the total flux transferred by the flare.   This inversion is similar in philosophy to the method used by \citet{Hori1997}, \citet{Reeves2002}, and \citet{Warren2002} to infer the energy input by fitting to one spectral band, and then applying the result to others.  The present case differs in that it infers a rate of flux transfer, and then produces energy through reconnection at that rate.

\begin{figure}[htbp]
\centerline{\includegraphics[width=3.6in]{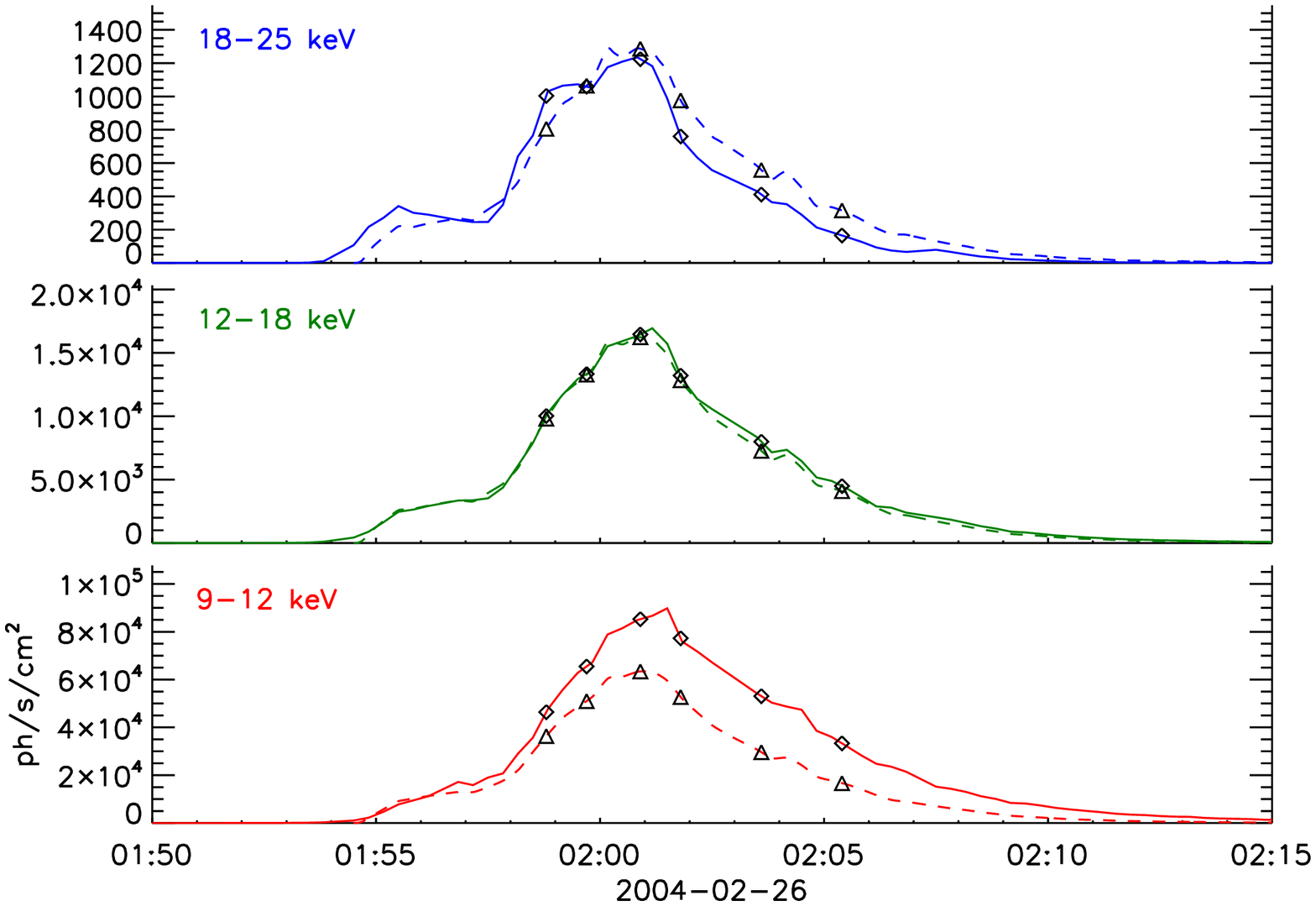}%
\includegraphics[width=3.0in]{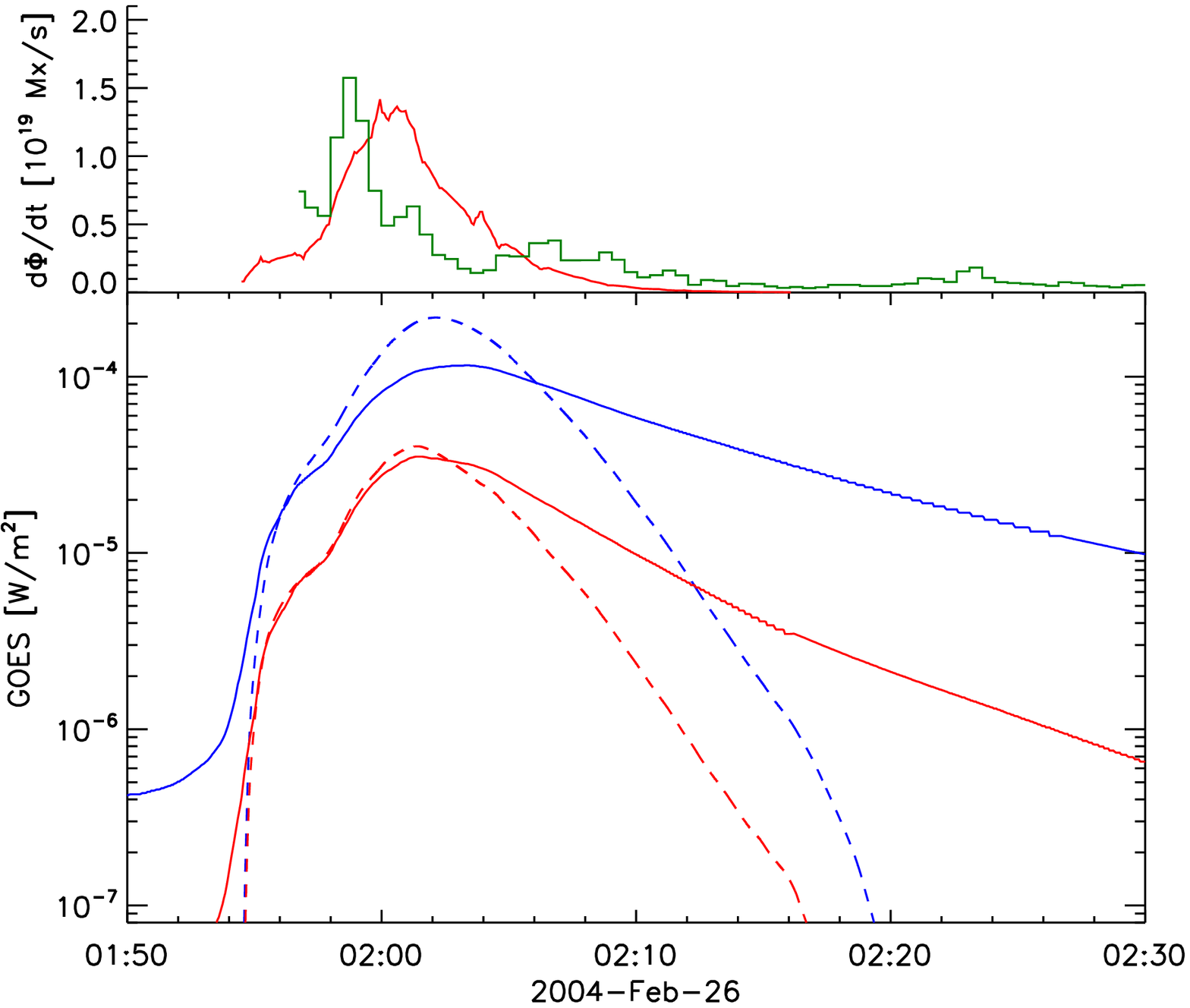}}
\caption{Light curves synthesized from the PREFT run (dashed) and observed (solid).  The left panel shows three HXR bands: $9$--$12$ keV (red), $12$--$18$ keV (green), and $18$--$25$ keV (blue).  Symbols correspond to the times of the spectra shown in \fig\ \ref{fig:specs}.  The lower right panel shows GOES curves from the $1$--$8$ \AA\ (blue) and $0.5$--$4$ \AA\ band (red).  The upper right panel shows flux transfer rates computed from flare ribbons (green) and HXR inversion (red), identical to those from \fig\ \ref{fig:synth}. }
	\label{fig:lc}
\end{figure}

The inverted flux transfer rate, $\dot{\Phi}_{\rm inv}(t)$, is next used to synthesize other observable bands, using eq\ (\ref{eq:conv}) with different response functions $R(T)$.  Figure \ref{fig:lc} shows, on the left, syntheses for three wider bands of thermal bremsstrahlung both above, below and covering the narrow band used for inversion.  The right panel shows the synthesis performed for both bands of GOES (solid).  Here we used the published response of GOES \citep{Garcia1994}, but multiply by an empirical correction factor of $0.5$, following observations by \citet{Hannah2008} and \paperI, of a discrepancy between GOES and RHESSI.  Each shows similarities with observations and telling differences we discuss below.  

\begin{figure}[htbp]
\centerline{\includegraphics[width=6.0in]{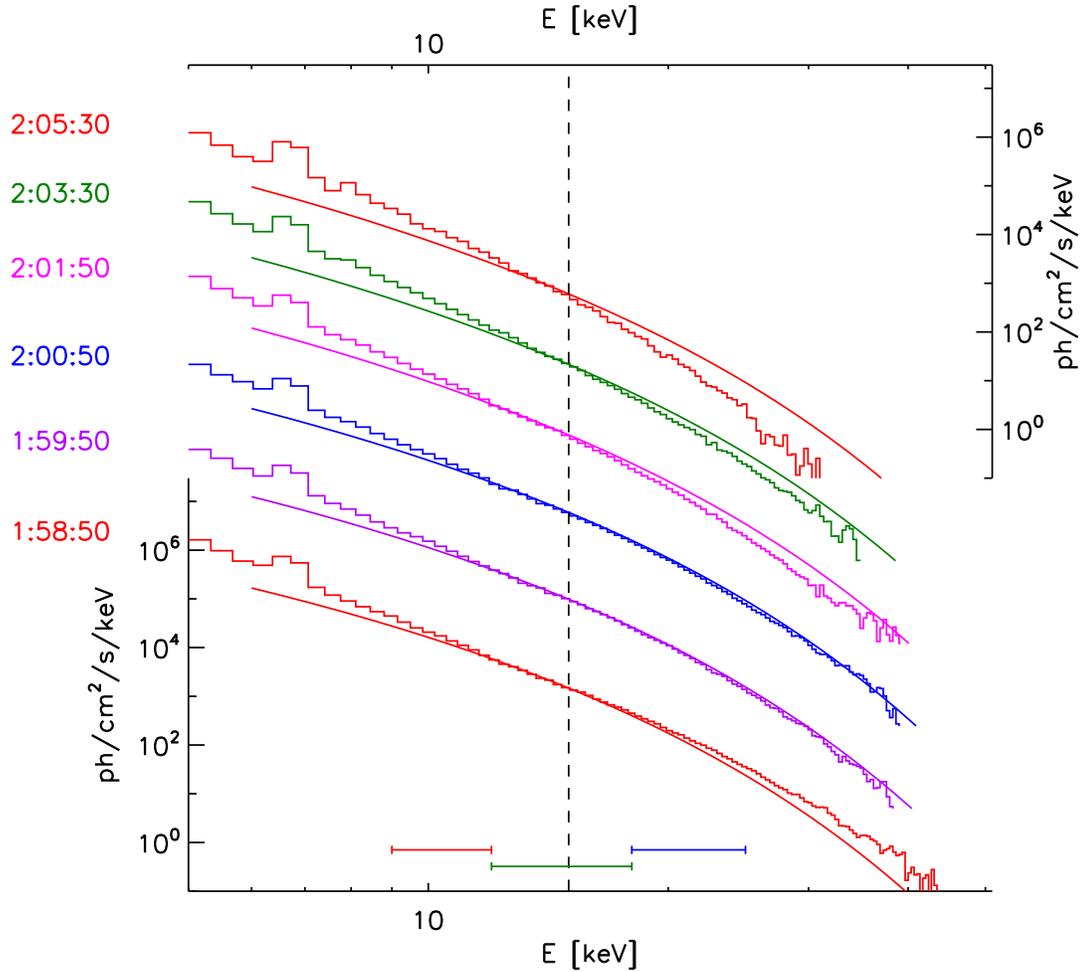}}
\caption{Spectra from six times through the peak of the flare, plotted in different colors, and displaced vertically for clarity.  Histograms show the data from RHESSI.  Smooth curves are synthesized from the PREFT run.  Colored bars along the bottom show the bands used to synthesize the light curves in \fig\ \ref{fig:lc}, and the vertical dashed line at $E=15$ keV is the center of the band used to derive the flux transfer rate $\dot{\Phi}_{\rm inv}$ in \fig\ \ref{fig:synth}.}
	\label{fig:specs}
\end{figure}

Complete X-ray spectra can be synthesized using this same procedure.  Temperature response function $R_i(T)$ gives the photon flux {\em per energy range} for spectral bin $i$, centered at photon energy $\varepsilon_i$, and having width 
$\Delta\varepsilon_i$.  As for \fig\ \ref{fig:synth} we consider the photon flux at 1 AU from thermal brems{\-}strah{\-}lung, but now divide by $\Delta\varepsilon_i$.  Figure \ref{fig:specs} shows the synthetic spectra obtained at six different times of the flare.  These are compared to the spectra measured by RHESSI at the same times.  The spectrum from 2:00:50 (blue) was used to help select the three initial parameters otherwise unconstrained by direct observation: 
$L_0=70$ Mm, $T_{0,{\rm max}}=7.5$ MK, and $\Delta\theta=110^{\circ}$.  It is noteworthy, nevertheless, that a TFT run from reasonable initial conditions, and lacking any non-thermal electrons, produces synthetic spectra matching observations in both shape and amplitude.  This agreement demonstrates that the super-hot material produced by SMSs can achieve sufficient emission measure to match observations.

\section{Synthesizing an HXR image}

Having synthesized disk-integrated light curves we now synthesize an HXR image.  We consider first a single tube whose flux is $\delta\psi=2\times10^{19}$ Mx, similar to the post flare loops visible in TRACE 171 \AA\ images such as \fig\ \ref{fig:trace}.  This assigns, for the first time, a cross-sectional area of $\delta\psi/B=10^{17}\,{\rm cm}^2$ to the PREFT run.  The resulting cell volumes are multiplied by $n_e^2$ and the temperature response $R(T)$ for thermal bremsstrahlung over photon energies spanning the 12--25 keV band \citep{Rybicki1979}, to synthesize spatially resolved emission at every time.

\begin{figure}[htbp]
\centerline{\includegraphics[width=6.5in]{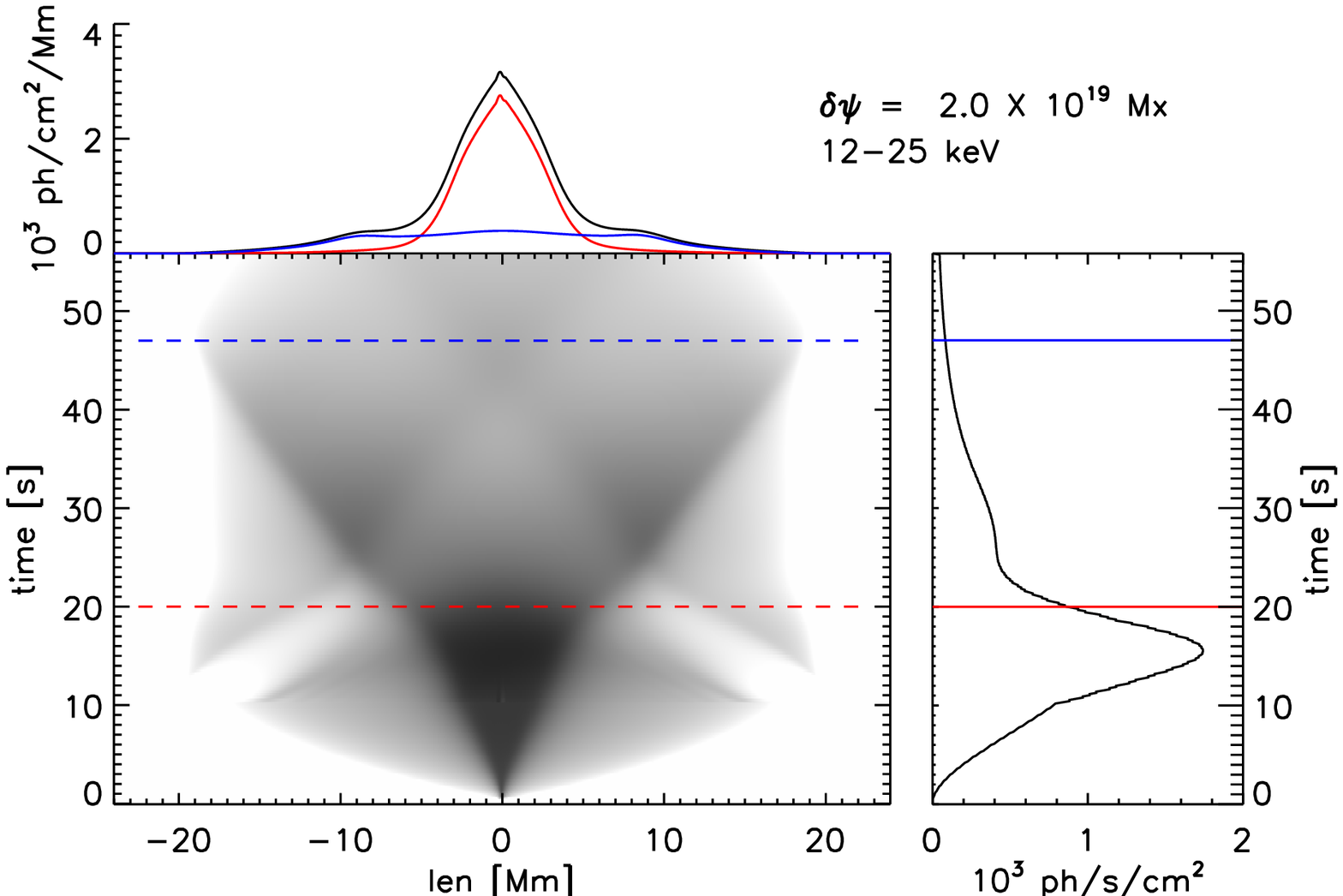}}
\caption{Time {\em vs}.\ space stack plot of the synthesized thermal brehmsstrahlung for the 12--25 keV band from a $\delta\psi=2\times10^{19}$ flux tube.  The grey scale in the central panel shows the emission in inverse-logarithmic scale.  The spatial coordinate is length, measured from the mid-point.  The right panel shows the spatially integrated light curve, 
$I_0(t)\delta\psi$.  Red and blue horizontal lines show the end times of distinct phases described in the text.  The top panel shows, in black, the result of time integrating the emission from each spatial coordinate.  The integral over just the first and second phases are plotted in red and blue respectively.}
	\label{fig:rh_stack}
\end{figure}

Figure \ref{fig:rh_stack} shows the result of this synthesis as a time {\em vs}.\ space stack plot.  The spatially integrated light curve, shown on the right panel, is $I_0(t)\delta\psi$, for the impulse response given in \eq\ (\ref{eq:I0}).  This has a single-peaked structure similar to that of the narrower band from \fig\ \ref{fig:synth}.  The spatial resolution makes it clear that the peak originates from the loop top.  The main peak, lasting until $t=20$ s (red line) is from the slow magnetosonic shocks during and shortly after retraction.  The slower decay ($20\,{\rm s}<t<47\,{\rm s}$, blue line) is from the phase where that source is re-compressed by evaporation.  These two phases combine to form a loop-top source in the time-integrated emission shown in the upper panel.  It is clear that the evaporation phase (blue) contributes negligibly to the loop-top source.

\begin{figure}[htbp]
\includegraphics[width=3.0in]{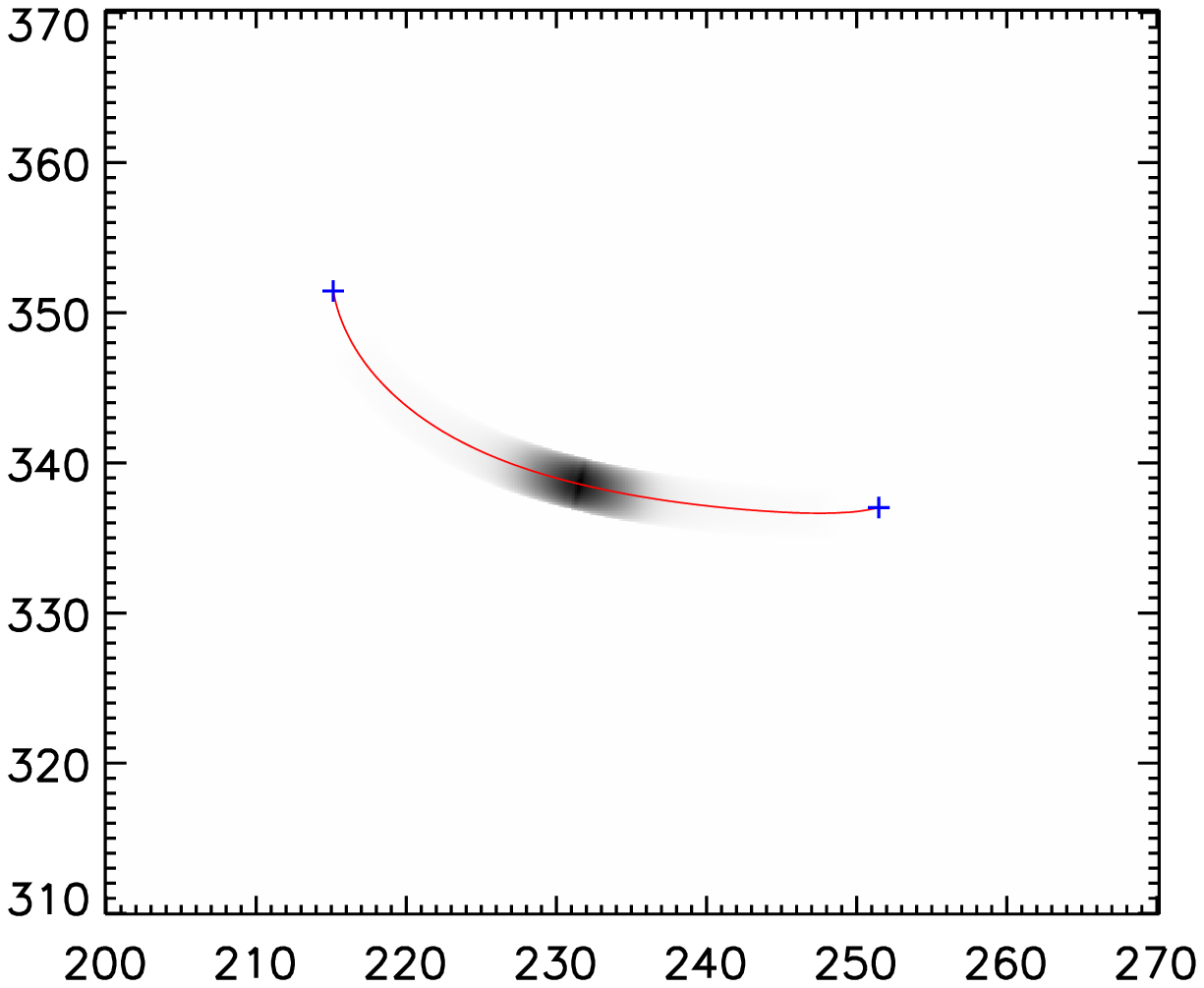}a%
\includegraphics[width=3.0in]{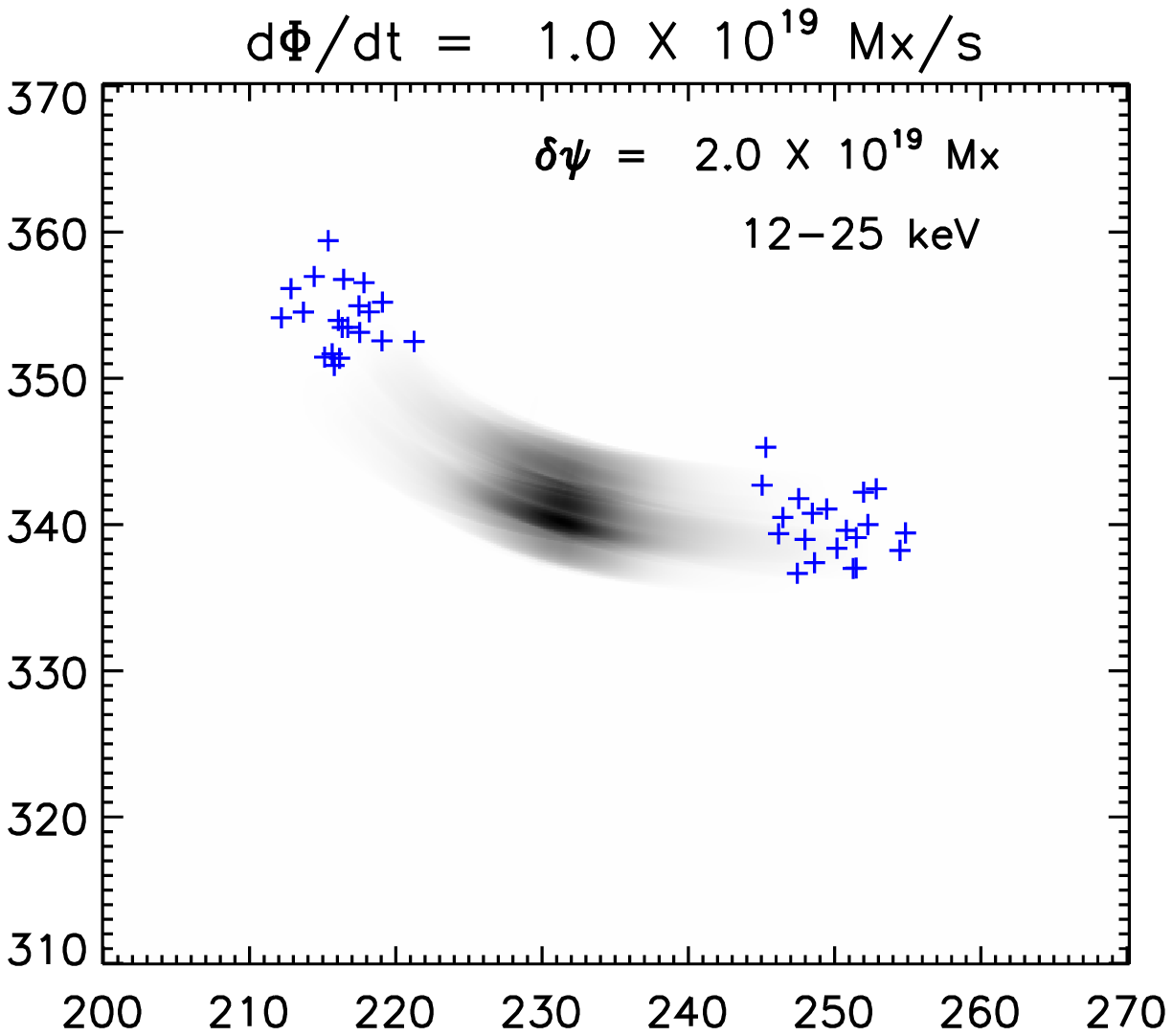}b
\includegraphics[width=3.0in]{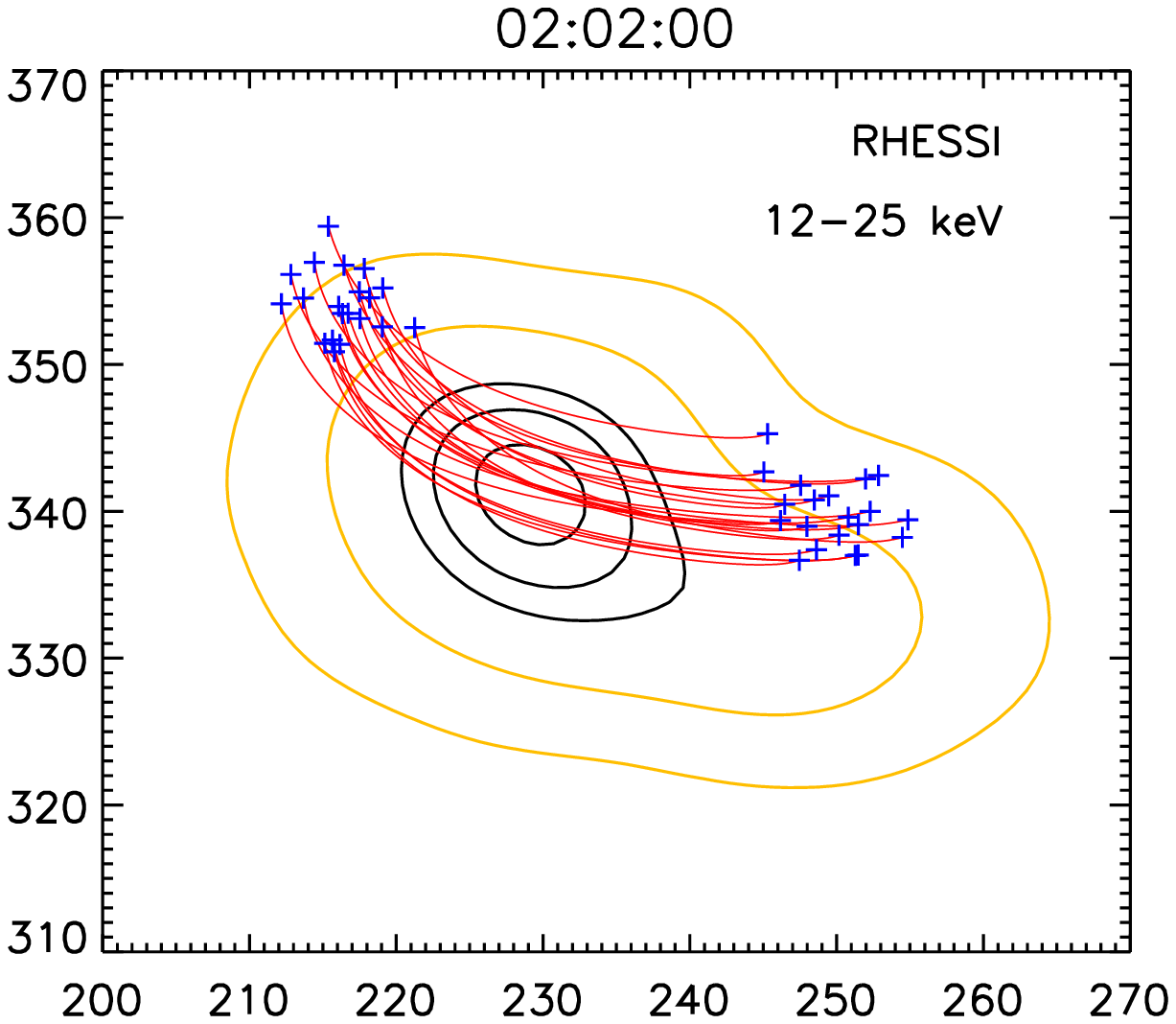}d%
\includegraphics[width=3.0in]{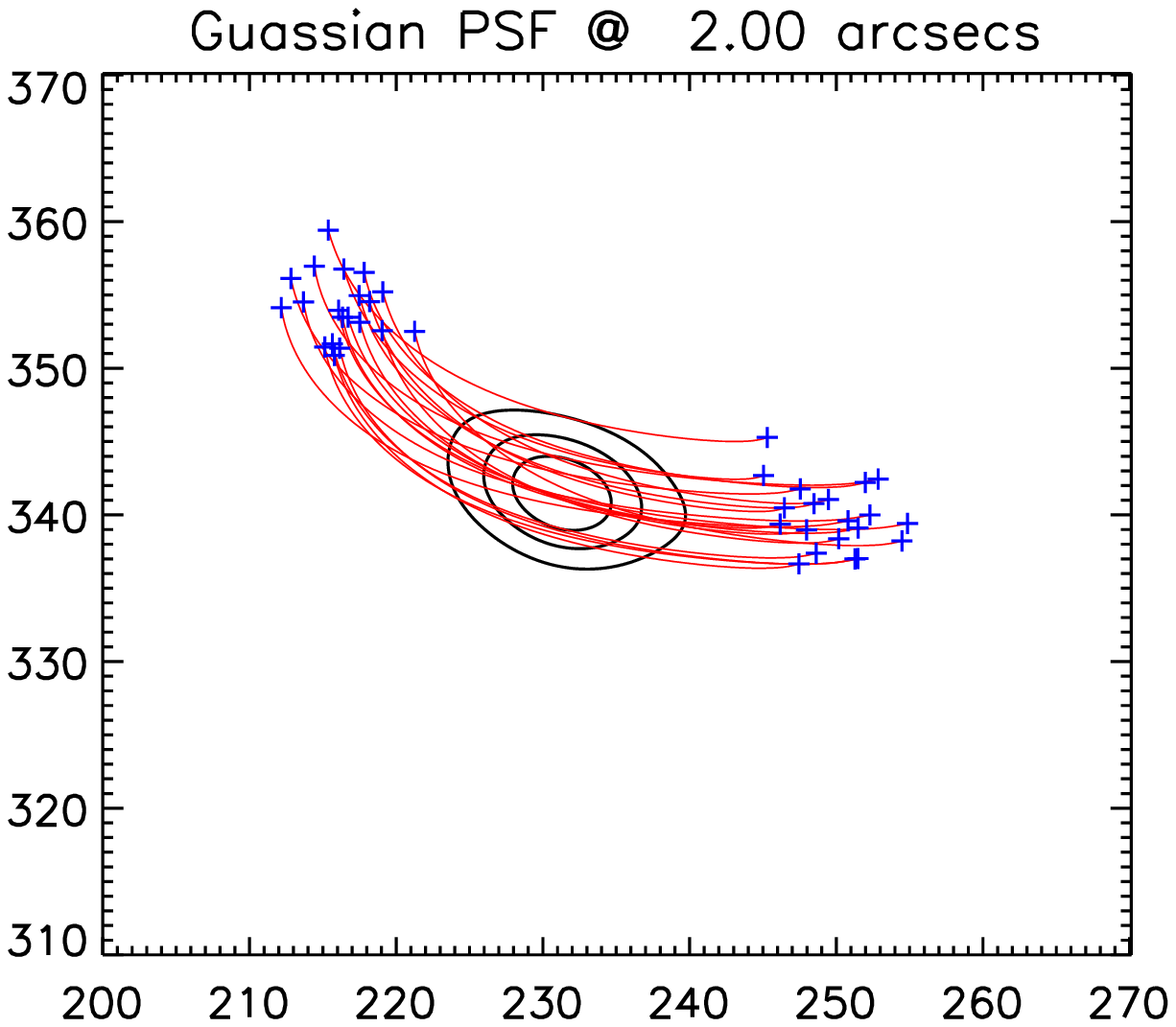}c
\caption{The synthesis of an HXR image depicted clockwise from upper left, a--d.  a. A single loop is formed by mapping the PREFT axis onto an extrapolated field line, shown as a red curve with blue crosses at its footpoints.  The plasma properties are passed through the 12-25 keV thermal bremsstrahlung response and integrated over a 20 s window.  The synthetic image is plotted in linear inverse grey scale.  b. 20 versions of the same loop, staggered in time by 2 s, and displaced slightly in space, shown in inverse grey scale, but normalized differently from a.  c. The image from b. is convolved with a Gaussian psf of 2'' width.  Black contours show 25\%, 50\% and 75\% of maximum.  d. The 12--25 keV RHESSI image computed using the CLEAN algorithm over a 20-sec integration shown as contours at 75\%, 85\% and 95\% of maximum (black) and  25\%, 50\% (orange).}
	\label{fig:rh_image}
\end{figure}

The spatially resolved emission just described is then mapped onto the plane-of-the-sky to produce an image.  The 2004-Feb-26 flare is on the disk, reasonably close to disk center, so any retraction motion would be extremely foreshortened.  Furthermore, the majority of super-hot emission occurs at the end of retraction ($5\,{\rm s}<t<10.2\,{\rm s}$) or after it has ceased.  We therefore map the entire time history of the PREFT simulation onto a static post-retraction loop.  We select one of the extrapolated field lines from \fig\ \ref{fig:loop_view} for this purpose.  The central point of the PREFT simulation is identified with the apex of the extrapolated loop.  Each grid point from the PREFT run is mapped to a point on the field line an equal distance from the apex.  The emission is then averaged over a 20 s to simulate the averaging performed in constructing a RHESSI image (see \fig\ \ref{fig:rh_image}a).

Our hypothesis is that reconnection achieved its overall reconnection rate $\dot{\Phi}$ by transferring a sequence of tubes each of flux $\delta\psi$.  The flux transfer rate at 2:02 was approximately $\dot{\Phi}=10^{19}$ Mx/sec, so one new tube is created every $\Delta t=\delta\psi/\dot{\Phi}=2$ s by reconnection.  Our full image thus consists of numerous different tubes at different stages of evolution following their own creation.  As an expedient measure we make 20 different copies of the single tube described above and displace them slightly over the plane of the sky.  The result is a set of 20 distinct tubes in various stages of evolution over 40 seconds, shown in \fig\ \ref{fig:rh_image}b.    We convolve this highly-resolved image with a Gaussian point-spread-function (PSF) of $2''$ width, to obtain the contours in  \fig\ \ref{fig:rh_image}c.  This compares favorably to the upper contours of the actual RHESSI image made using the CLEAN algorithm over the 12--15 keV band (\fig\ \ref{fig:rh_image}d, see \paperI).

\section{Discussion}

We have used a numerical TFT solution, from the PREFT code, to synthesize various observations from the flare of 2004-Feb-26.  Synthetic light curves (\fig\ \ref{fig:lc}), HXR spectra (\fig\ \ref{fig:specs}), and an image (\fig\ \ref{fig:rh_image}), all show reasonable agreement with counterparts from observation.
This agreement offers some level of support to the hypothesis, originally proposed in \paperI, that loop-top HXR sources are produced by slow magnetosonic shocks (SMSs) from Petschek-like reconnection.  The simulations confirm that SMSs create sufficiently high densities and temperatures to produce HXR spectra with shape and amplitude matching observations.  The predicted amplitude depends on the rate of magnetic reconnection, which is observed in proxy through the motion of flare ribbons.  Because the field-aligned compressive flows persist beyond the end of flux tube retraction, the dense emitting sources appear to rest at the top of the flare arcade, producing a compact, loop-top source.  These detailed comparisons are enabled by numerical simulation, but they serve to confirm that the analytic estimates used to make the original hypothesis in \paperI, were reasonable ones.

The PREFT simulations reveal that chromospheric evaporation plays a relatively minor role in producing the loop-top source.  Figure \ref{fig:rh_stack} shows that the second stage of the impulse-response, $20\,{\rm s}<t<47 {\rm s}$, accounts for a small portion of the emission in 12--25 keV.  This stage represents the re-compression of the loop-top SMSs by evaporation, which creates a broad, dense, cooler concentration over most of the loop.  It stands to reason that evaporated material will be cooler than the source of conduction driving the evaporation, and our simulation confirms this.  For this reason the blue curve peaks at lower temperature than the red curve in the top panel of \fig\ \ref{fig:DEM}.  This points to the chief problem with invoking evaporation in the role of creating the high coronal density required for loop-top sources --- even more problematic than the time delay ($\sim20$ s in our model).

The TFT model requires, as input, five parameters describing the flux tubes created through magnetic reconnection.  Two of these, the field strength and the final retracted length, were constrained using post-flare observation.  In this work we derive the remaining three parameters by fitting the RHESSI spectra from the peak time.  One of these, the length of the tube just after reconnection but before retraction, is found to be $\Delta L=24.8$ Mm longer than the length after retraction.  This sets the free energy which ultimately powers the flare, and is consistent with a CS extending about 
$h=\Delta L{\rm cot}(\Delta\theta/4)/2=24$ Mm above the flare arcade.  This fairly short CS seems consistent with a compact flare from which there was no evidence of eruption \citep{Wang2007}.  The downward viewing angle would make it even more difficult to observe this CS than those from eruptive limb flares.  Nor do we expect 171 \AA\ images to reveal the CS.

The shear angle at the current sheet is found to be $\Delta\theta=110^{\circ}$. This means the reconnection occurs with a guide-field component, $B_x=\cot(\Delta\theta/2)B_z$, roughly 70\% as large as the reconnection component.  A significant guide field was predicted in \paperI\ for this particular AR, where new flux emerged roughly parallel to existing flux.  They used a magnetic model to predict $\Delta\theta\simeq70^{\circ}$, however, a shear angle so small would produce SMSs with density and temperature too low to match observation.  Indeed, \paperI\ required $\Delta\theta\simeq100^{\circ}$ to obtain agreement from their analytic estimates based on RH conditions.  Thermal conduction yields lower temperatures and higher densities than predicted by RH, so our model required a still larger shear angle.

The most surprising requirement for spectral fitting is that the pre-retraction flux tube must have an apex temperature of $T_0=7.5$ MK.  This temperature is required to produce an equilibrium flux tube with an initial density of 
$n_e=1.7\times10^{10}\,{\rm cm}^{-3}$.   \paperI\ had anticipated the need for a large initial density, although their analytic estimates led to a value 5 times higher still.  High densities are required primarily to produce, through compression at SMSs, the high densities observed in the loop-top source.  They also set an Alfv\'en speed sufficiently low to keep the HXR spectra as soft as it is observed to be.  Other lines of investigation have found that fitting observed emission measures requires initial densities as high, or higher, than ours \citep{Reep2016}.  

It must be borne in mind that our initial condition is that of field lines inside a current sheet which have not yet undergone reconnection --- it is not the conditions of the ambient, pre-flare loop.  It is possible that flux within the current sheet has experienced some heating and ``densification'' prior to its actual reconnection \citep[offers one possible mechanism]{Scott2013}.  Another possibility is that reconnection of other flux, during the preceding phase of the same flare, might have supplied energy across field lines and driven evaporation into that flux tube.  That earlier reconnection would have occurred at lower density, thus producing  harder HXR spectra with lower emission.  Such a progression to harder spectra as the emission rises is what is actually observed in \figs\ \ref{fig:lc} and \ref{fig:specs}.  During the earliest phase of this process the pre-retraction density might have been low enough to prevent coronal thermalization, thus admitting a non-thermal electron population to reach the footpoints.  This is what does actually occur in this flare, as reported in \paperI\ and seen in \fig\ \ref{fig:thist} here.

\citet{Veronig2005} observed a different loop-top source without footpoints, whose apex density they estimated to be $n_e\simeq1.7\times10^{11}\,{\rm cm}^{-3}$ at the flare's peak.  They found in pre-flare HXR emission and Nobeyama 17 GHZ radio images, evidence that loops had been enhanced to densities $n_e\simeq5\times 10^{10}\,{\rm cm}^{-3}$ even before the flare began.  Their explanation, anticipating the foregoing, was that pre-flare activity had driven chromospheric evaporation enhancing the density in those loops.  We believe it is possible that evaporation has played this role in our flare as well, raising pre-reconnection density, but is {\em not} responsible for the factor of four increase {\em during} reconnection.

Having chosen parameters for a single flux tube, we use its time-evolving thermal bremsstrahlung emission to synthesize HXR spectra over the course of the flare.  The synthesized spectra from 1:59:50 and 2:00:50 (around the flare's peak) match the observations nicely for energies above 10 keV.  The model's deficiency at lower energies may stem from its neglect of emission other than bremsstrahlung, and possibly from its failure to account for emission from surrounding flux awaiting reconnection --- the same flux we assume to be at $7.5$ MK.   (The process of converting RHESSI's counts to photon spectra, as we have done, is also known to be less reliable below 10 keV.)   

Since our model lacks non-thermal electrons, the synthesized spectrum is due to thermal bremsstrahlung alone -- it is a thermal model.  It is not, however, a model with one or two temperatures, but is derived from a DEM undergoing self-consistent time-evolution as shown in \fig\ \ref{fig:DEM}.  Because this evolution is relatively rapid, $\sim 10$ s, the synthesized spectra have very similar structure over the minutes-long flare (compare the solid curves of \fig\ \ref{fig:specs}).  The observed spectrum, in contrast, softens steadily during the flare, so the model matches well for only a minute.  Matching the slower softening would require a series of different flux tubes whose properties change over the course of the flare.  Such evolution is natural, since we expect ongoing reconnection to decrease the shear angle, $\Delta\theta$, and field strength, $B$, of the current sheet, and to increase the initial length of reconnected tubes, $L_0$.  Each of these is likely to soften the synthetic spectrum.  Such a model will clearly be more complex than the one presented here, and must be undertaken in the future.

Our model synthesizes observables by convolving the simulated emission with a time-varying reconnection rate $\dot{\Phi}(t)$.  The reconnection rate can be measured from the motion of flare ribbons \citep{Forbes1984,Poletto1986,Qiu2002}, and we do so using 171\AA\ images to obtain $\dot{\Phi}_{\rm rib}(t)$.  We find, however, that observations are better reproduced using an inverted curve, $\dot{\Phi}_{\rm inv}(t)$, which has a similar profile, peaks at a similar level, $\dot{\Phi}\simeq1.6\times10^{19}$ Mx/s, but two minutes later than 
$\dot{\Phi}_{\rm rib}(t)$.  This discrepancy may point to a delay between the forging of a new, reconnected flux tube, and its retraction under tension.  Alternatively, it may result from the complex chromospheric physics underlying the formation of the flare ribbons.  The peak reconnection rate is equivalent to 160 gigavolts, and not dissimilar to values observed in other large flares \citep{Qiu2002,Qiu2004,Qiu2010}.   This value represents the rate at which new flux tubes are produced by reconnection, but we do not assume it is related in any simple, direct way to the local reconnection electric field as previous investigations have assumed \citep{Forbes1984,Poletto1986,Qiu2002,Qiu2004,Isobe2005}.  Instead, our work tacitly assumes this local electric field is simply very large, and thus forges a new flux tube rapidly enough to produce little global effect.  This might proceed in a quasi-steady manner or more sporadically to produce distinct flux tubes.  The transfer rate $\dot{\Phi}(t)$ applies equally to either scenario, and produces exactly the same synthetic light curves.

Reconnection in the model, as in observations, transfers a net flux $\Delta\Phi\simeq6\times10^{21}$ Mx over the course of the flare.  This is 30\% of the total flux in the AR as reported by \paperI.  Multiplying this flux by the magnetic energy released by a single flux unit gives a total energy release of $\Delta W_m=2.4\times10^{32}$ ergs, as shown in \fig\ \ref{fig:erg_plot}.  This energy is converted initially to bulk kinetic energy as the flux tubes retract.  Only 25\%, roughly $6\times10^{31}$ ergs, remains within the flux tube to produce the effects of the flare.  This compares favorably to $4\times10^{31}$ ergs, found by \paperI\ using GOES light curves and observed loops to compute radiative and conductive losses during the course of the flare.  Our model eliminates the remaining 75\% artificially when the tube reaches is final length.  This elimination is meant to represent those processes which stop the tube, including the emission of fast magnetosonic waves or even formation of a fast magnetosonic termination shock.  \citet{Longcope2012} argued, using a crude two-dimensional MHD calculation, that the process of eliminating a CS by reconnection could produce fast magnetosonic waves carrying roughly this fraction of the released energy.

The present work has applied the TFT model to a single, well-observed flare in order to make a compelling case for the viability of SMSs as the explanation for a loop-top, super-hot HXR source.  We believe, however, that the model is more widely applicable.  Magnetic energy is believed to power all solar flares, and must be released primarily by shortening field lines which have stored the energy.  This shortening will naturally compress the plasma contained there.  If the retraction occurs rapidly enough the compression will occur as shocks (SMSs) which are the fundamental elements in the Petschek reconnection model.  This rapid compression produces high temperature and high density at the same time.  These are the two key elements of observed loop-top sources, and models lacking SMSs must invoke separate mechanisms for each element.  The SMSs thus seem fairly natural in the role of loop-top sources, and the foregoing has shown them to be viable.  Of course, applying them to cases with non-thermal particles will require a generalization of the strict fluid approach adopted here.  That will need to be pursued in future work.

\acknowledgements

This work was supported partly by a grant from NASA's Heliophysics Supporting Research (HSR) program and partly by a grant from NSF/AGS's Research Experiences for Undergraduates (REU) program.  The authors thank Amir Caspi for discussions and Marina Battaglia for advising on the analysis of RHESSI data.


\end{document}